\documentclass[a4paper,fleqn,usenatbib]{mnras}
\usepackage{txfonts}
\usepackage[T1]{fontenc}
\usepackage{ae,aecompl}
\usepackage{ifpdf}
\usepackage{hyperref}
\usepackage{graphicx}

\newcommand{\gamete}{\textsc{GAMETE/QSOdust}}
\newcommand{\cloudy}{Cloudy}

\title[BH seeds signatures]{Chasing the observational signatures of seed black holes at $z>7$: candidate observability}

\author[Valiante et al.]{Rosa Valiante$^1$ \thanks{E-mail: rosa.valiante@oa-roma.inaf.it},
Raffaella Schneider$^{1,2}$, Luca Zappacosta$^1$, Luca Graziani$^1$, \and
Edwige Pezzulli$^{1,2,3}$ and Marta Volonteri$^{4}$ \\
$^1$ INAF-Osservatorio Astronomico di Roma, via di Frascati 33, 00078, Monteporzio Catone, 
Italy\\
$^2$ Dipartimento di Fisica, Universit\'a di Roma “La Sapienza”, P.le Aldo Moro 2, 00185, Roma, Italy\\
$^{3}$INFN, Sezione di Roma I, P.le Aldo Moro 2, I-00185 Roma, Italy\\
$^4$ CNRS, UMR 7095, Institut d'Astrophysique de Paris, F-75014, Paris, France
}

\date{Accepted . Received }
\pubyear{2017}

\begin{document}
\label{firstpage}
\pagerange{\pageref{firstpage}--\pageref{lastpage}} 
\maketitle

\begin{abstract}
Observing the light emitted by the first accreting black holes (BHs) would 
{dramatically}
improve our understanding of the formation of quasars at $z>6$, 
possibly unveiling the nature of their supermassive black hole (SMBH) seeds.
In previous works we explored the relative role of the two main competing BH seed
formation channels, Population~III remnants (low-mass seeds) and direct collapse BHs (high-mass seeds), 
investigating the properties of their host galaxies in a cosmological context.
Building on this analysis, we predict here the spectral energy distribution and observational 
features of low- and high-mass BH seeds selected among the progenitors of a $z\sim 6$ 
SMBH. We derive the processed emission from both accreting BHs and stars by using the photo-ionization 
code \cloudy, accounting for the evolution of metallicity and 
dust-to-gas mass ratio in the interstellar medium of the host galaxies, as predicted by 
the cosmological data-constrained model \gamete.
{We show how future missions like JWST and ATHENA will 
be able to detect the light coming from SMBH progenitors already at $z\sim 16$. 
We build upon previous complementary studies and propose
a method based on the combined analysis of near infrared (NIR) colors, 
IR excess (IRX) and UV continuum slopes (i.e. color-color and IRX-$\beta$ diagrams) to 
distinguish growing seed BH host galaxies from starburst-dominated systems in JWST surveys.
}
Sources selected through this criterion would be the best target for follow-up X-ray observations.
\end{abstract}


\section{Introduction}

{Recent theories of BH formation predict at least two viable mechanisms explaining the formation of accreting 
SMBHs of $10^9-10^{10} \, M_\odot$, which are observed already at redshift $z\sim 7$: 
the rapid growth of Population~III (Pop~III) remnant BHs ($\sim 100 \, M_\odot$)
\citep[e.g.][]{MadauRees01,Heger03,Yoshida08,Latif13b,Hirano15}, 
driven by the accretion of gas at rates exceeding the Eddington limit 
or the formation of more massive BH seeds, 
the so-called direct collapse BHs (DCBHs), of $(10^5-10^6) \, M_\odot$ 
\citep[e.g.][]{BL03,LodatoNatarajan06,
LodatoNatarajan07, Begelman06, Volonteri08,IO12, 
Ferrara14, Agarwal14, Inayoshi15}.
We refer the readers to a number of dedicated reviews for more details on different seed formation and growth mechanisms 
\citep[][and references therein]{Volonteri10,Natarajan11,LatifFerrara16,JohnsonHaardt16,Gallerani17,V17a}.
}


{Several semi-analytic models have been devoted to study the formation of SMBHs, exploring the role
of low-mass or high-mass seeds \citep[e.g.][and references therein]{Volonteri03,Volonteri05,Volonteri06,
Madau04,Haiman04,TH09,VSD15}. \citet{Petri12} presented the first semi-analytic model for the formation 
of high redshift SMBHs exploring the combined action of low- ($100 \, M_\odot$) and high-mass ($10^5 \, M_\odot$) 
seed BHs. In all those works, however, seeds were not planted according to the environmental conditions and the 
chemical evolution of the interstellar medium (ISM) of the host galaxies was not considered.
}

In a previous work \citep[][hereafter V16]{V16}, we used a cosmological, data-constrained model,
\gamete \, to investigate the relative role of low-mass (light) and high-mass (heavy) BH seed in 
the formation of $z>6$ SMBHs, like the one powering the quasar SDSS J1148+5251. 
{\gamete \, consistently follows the formation and evolution of nuclear BHs and their host galaxies along 
different merger histories of the parent DM halo. While many semi-analytical models and simulations follow BH growth,
star formation and metals, \gamete \, is the only model that traces accurately dust evolution as well.} 

In V16 we showed that in a Eddington-limited BH accretion scenario, despite their low occurrence (in just 
few halos, 3-30 depending on the merger history), the formation of 
DCBHs is required to grow a SMBH with $M_{\rm BH}>10^9 \, M_\odot$.
Moreover, in \citet{P16} we showed that a mildly super-critical growth (less than 20 times 
the Eddington rate) onto light BH seeds of $100 \, M_\odot$ is required to reproduce the mass 
of SDSS J1148+5251 if heavy seeds are not considered in our model.

Despite intensive efforts, no observational signatures of high redshift SMBHs progenitors have 
been found in current surveys\citep[see e.g.][for a recent review]{ReinesComastri16}.
To date, no clear AGN signatures at $z>6$ have been found even in the deepest X-ray 
observations \citep{Willott11, Treister13, Weigel15, Vito16, Cappelluti16, Luo17}. 
It should be noted though, that two faint AGN candidates, with photometric redshift $z=6.6$ and $z=9.7$, 
have been selected in the CANDELS GOODS-S field \citep{Giallongo15}.
{Although the redshift determination is still uncertain (\citealt{Georgakakis15, Cappelluti16, Luo17}), 
these sources have been proposed as possible DCBH hosts by \citet{Pacucci16}. }

A clear example of the difficulties in interpreting observations at high redshift is the controversial 
nature of the metal poor component of the bright $\rm Ly\alpha$ emitter CR7 observed by 
\citet{Matthee15}, \citet{Sobral15}, and \citet{Bowler16} at redshift $z\sim 6.6$. 
This is suggested to be either a DCBH 
\citep[][]{Pallottini15, Hartwig15, Agarwal16b, Smith16, Smidt16, Agarwal17a, Pacucci17} or a Pop~III 
stars host candidate \citep{Sobral15, Visbal16, Dijkstra16}, or none of the two \citet{Bowler16}.

If faint AGNs at high redshift are powered by accreting DCBHs,  the non detection of AGN candidates 
in current surveys may be explained by the peculiar and rare environmental conditions required 
for DCBHs formation (\citeauthor{Hosokawa12} \citeyear{Hosokawa12, Hosokawa13}; 
\citealt{Inayoshi14}; \citealt*{IH14}; \citealt{Yue14, Dijkstra14, Sugimura15, Habouzit16hSAM, 
Chon16, V16}).
On the other hand, the current lack of detection in the X-ray band 
could be due to short and intermittent super-Eddington accretion events on Pop~III remnants, 
resulting in a very low active BH fraction $(\sim 1\%)$ at $z > 7$ \citep{P17a}.

Theoretical models of the expected spectral energy distribution (SED) of AGNs at $z>4-5$  may provide 
a powerful tool to support future observations. Indeed, with the upcoming
\textit{James Webb Space Telescope} JWST and \textit{Advanced Telescope for High Energy Astrophysics} (ATHENA, in the next decade) missions we may be able to detect the signatures of active 
galaxies up to very high redshifts. \citet{Pacucci15} modeled the SED of a growing DCBH with initial mass of $10^5 \, M_\odot$ accreting 
gas in a metal-free isolated DM host halo.
They use the code Cloudy \citep{Ferland13} to process the intrinsic (optical/UV/X-ray) emission, 
suggesting that the resulting infrared (IR) emission around $\sim 1~\mu$m and the $[0.1-10]$ keV band 
X-ray emission may be detected by JWST and ATHENA.
Subsequently, \citet{Natarajan17} extended this study to what they call  {\it obese black hole galaxies} 
(OBGs), a class of transient objects predicted by \citet{Agarwal13} that originate from the merging of 
star forming galaxies with a neighbor satellite DCBH host. 
The authors presented a color-color selection criteria aimed at identifying OBGs with predicted 
magnitudes $\rm M_{AB}<25$, that should be detected by the JWST  instrument MIRI.  
They compared the SEDs of OBGs and of growing Pop~III remnant BHs at redshift $z=9$, concluding 
that JWST could be able to discriminate among the two seed formation channels. 
In addition, \citet{Volonteri17} presented a population synthesis model for high redshift 
BHs, AGN and galaxies, based on empirical relations, proposing a color-color selection
in JWST photometric bands, F1280W-F2100W vs F200W-F1280W, to discriminate starburst 
dominated galaxies and AGN in future surveys.

The studies mentioned above introduced promising methods to search for AGN and seed BHs candidates in 
future surveys, setting the (theoretical) groundwork for the upcoming high redshift multi-wavelength 
observations. 
{However, the systems analyzed in \citet{Natarajan17} and \citet{Volonteri17} are not 
explicitely related to the formation of a SMBH with the mass and accretion rate sufficient to explain the luminous quasars at $z\sim 6$ and do not grow in its environment. 
}

In \citet[][hereafter paperI]{V17b} 
we presented a statistical analysis of the properties of ``isolated'' galaxies that host
light (low-mass) or heavy (high-mass) BH seeds\footnote{Starting from the epoch of the BH seed 
formation, we consider a system (BH and its host galaxy) to evolve in ``isolation'' until a minor 
or major galaxy merger takes place.}. 
We named these systems (BH $+$ host galaxy) \textit{isolated light seeds} (ILS) and 
\textit{isolated heavy seeds} (IHS). Following V16, ILS and IHS have been selected among the 
progenitors of SDSS J1148+5251.
{In paperI we pointed out that at $z>10$ the majority of heavy ($\sim 98\%$) and light ($80\%$) 
seeds evolve in isolation
and that these two} 
populations show different properties: IHS have BH accretion rate higher than ILS;
IHS host galaxies show a factor of 5-10 lower stellar mass and metallicity as well as a less efficient
star formation history. These differences becomes statistically negligible at $z\leq 10$, when the fraction
of IHS (ILS) drops to less than $\sim 2~(20)\%$ as a consequence of merger events.
At these later epochs, any trace of the BH seed origin and of its birth environment is lost.
Thus, candidate systems at $z>10$ are the best targets to discriminate among the
two main seed formation channels.

Here we compute the SED of IHS and ILS, using the sample presented in paperI.
Our aim is to establish the feasibility of: 
({\it i}) detecting high-$z$ faint progenitors of the first SMBHs, ({\it ii}) observationally discriminating 
light and heavy BH seeds. 
Similarly to \citet{Pacucci15} and \citet{Natarajan17}, we use Cloudy
to model the SEDs of IHS and ILS. 
The main difference of our approach is that we analyze systems (BHs-host galaxies) 
selected among the ancestors population of an observed SMBH, 
{along its cosmological evolution. In other words,} our approach has the advantage of following the evolution of light and heavy seed progenitors along 
the same hierarchical history (i.e. the two seed BHs formation channels are not mutually exclusive), 
starting from the epoch at which the first stars form ($z\sim 24$), 
down to $z\sim 6$, when the observed quasar is eventually assembled. 
The properties of the seed BH population are closely related to the evolution of their host 
galaxies: mass, number, redshift distribution and growth history are regulated by the build up
of the UV radiation field, by the metals and dust pollution of the 
ISM and intergalactic medium (IGM) and by the effect 
of stellar and AGN-driven winds.
Our approach allows to asses the possibility of observationally distinguishing the two different 
populations of seed BHs.

The paper is organized as follows. In Section 2 we present the model for the SEDs. 
In Section 3 we discuss in details two accreting heavy and two accreting light seed BH prototypes, 
inspecting their predicted SEDs. Finally, we discuss our results drawing the conclusions of the 
study in Sections 4 and 5.

\section{Summary of the model}\label{sec:model}

Here we summarize the main features of \gamete, a semi-analytic, data-constrained 
model developed to study the formation and evolution of high-redshift quasars and 
their host galaxies in a cosmological framework.
{
We refer the reader to \citet[][and paperI, for further details on \gamete]{V11,V14,V16}
\footnote{The model \textsc{gamete} (GAlaxy MErger Tree and Evolution) was originally 
conceived to study the formation and evolution of local, Milky-Way like galaxies, in a 
cosmological context \citep[][]{SS07, SS08}}.
}

\gamete \, successfully reproduced the observed properties of a sample of quasars 
at $z>5$ \citep[][]{V14}.
Following V16, we focus our analysis 
on the progenitor BHs of the $(2-6)\times 10^9 \, M_\odot$ SMBH which powers 
SDSS J1148+5251, one of the best studied luminous quasars at $z>6$. A summary of the
properties of SDSS J1148+5251 inferred from the observations can be found in \citet{V11,V14}.

\paragraph*{Dark matter halo merger trees.}
Different hierarchical merger histories (merger trees) of the $10^{13} \, M_\odot$ host dark matter (DM) halo
have been reconstructed by using a Monte Carlo algorithm following the Extended Press-Schechter 
theory. The DM halo resolution mass at $z>14$ is extended to halos with virial temperature in the range
$1200$ K $\leq T_{\rm vir}<10^4$ K (i.e. mini-halos). These mini-halos dominate the halo mass spectrum
at $z\gtrsim 17$ and are expected to host the first generation of stars (Pop~III stars). 
At $z>14$ instead the merger tree halo population is dominated by Ly$\alpha$ halos,
namely DM halos with $T_{\rm vir}\geq 10^4$ K (see V16 for details).

\paragraph*{BH growth.}
The growth of the BH is driven by gas accretion and mergers. 
The BH accretion rate is described by the Bondi-Hoyle-Lyttleton formula 
re-scaled by a factor $\alpha_{\rm BH}=50$ (to match the BH mass of 
SDSS J1148+5251)\footnote{$\alpha_{\rm BH}$ is a free parameter required in semi-analytic 
models and numerical simulations to account for the higher central densities around 
accreting BHs \citep[e.g.][]{DiMatteo05}.} and can not exceed the Eddington limit \citep[see][for details]{V14}.

\paragraph*{Star formation and ISM chemical evolution.}
In each progenitor galaxy, stars are assumed to form following a series of quiescent and/or 
major merger-enhanced bursts\footnote{A major merger is defined as the coalescence of two DM halos with
mass ratio (less massive over the most massive) $\mu > 1/4$}. 
Following each star formation episode, the production of metals and dust by Asymptotic Giant Branch 
(AGB) stars and Supernovae (SNe) is regulated by the stellar lifetimes. In addition, we include 
physical prescriptions for dust processing in a two-phase ISM: destruction by SN shocks in the hot, 
diffuse medium and grain growth in cold, dense molecular clouds \citep[see][for details]{V14, deBen14}. 

\paragraph*{Mechanical feedback.}
We also include energy-driven winds, triggered by SNe and accreting BHs. 
The metal and dust rich gas outflows, launched on galaxy-scales, are responsible for polluting the IGM. 
This mechanical feedback is described assuming that $0.2~(0.25)\%$ of the energy released by SN 
explosions (BH accretion) is used to accelerate the ISM gas \citep[see][and V16 for details]{V12}. 
As observed in SDSS J1148+5251 \citep[][]{Maiolino12, Cicone15}, in our model the evolution history 
of BHs and their host galaxies is mainly regulated by AGN feedback, with SN-driven winds having only 
a minor effect.

\subsection{First stars and BH seeds formation}
As discussed in V16, the number and nature of stars and BH seeds is determined by the 
combined effect of chemical and radiative feedback.

The level of chemical enrichment determines the duration of the Pop~III star formation era. 
Pop~III stars form out of metal poor gas when the ISM metallicity is lower than the adopted critical 
threshold $Z_{\rm cr}\sim 10^{-4} \, Z_\odot$ \citep[][]{Schneider02, Schneider03, Schneider12}. 
The Pop~III initial mass function (IMF) is assumed to be a Larson IMF \citep[][]{Larson98} with a 
characteristic mass $m_{\rm ch}=20 \, M_\odot$ (top-heavy IMF), whereas for Pop~II we adopt 
$m_{\rm ch}=0.35 \, M_\odot$ (standard IMF).

In addition, the level of intensity of $\rm H_2$ photo-dissociating radiation, to
which halos are exposed, determines whether stars (in mini-halos) or DCBHs 
(in Ly$\alpha$ cooling halos) form and regulates the star formation efficiency in mini-halos 
(according to the halo virial temperature, redshift and metallicity, see Appendix A in V16 and \citealt{deBennassuti17}). 

When the star formation efficiency is low, we perform a stochastic sampling of the intrinsic Pop~III 
top-heavy IMF, randomly selecting stars in the mass range $[10-300] \, M_\odot$ until the total stellar 
mass formed in the burst is reached. Light BH seeds
are formed by the collapse of Pop~III stars with masses in the range 
$[40-140] \, M_\odot$ and $[260-300] \, M_\odot$\footnote{We assume that only the most massive BH among 
the Pop~III remnants formed in each burst settles in the center of the halo accreting gas from its 
surrounding.}.

In Ly$\alpha-$cooling halos, heavy BH seeds ($10^5 \, M_\odot$) can form 
when $Z<Z_{\rm cr}$ and the cumulative Lyman Werner (LW) emission (from all galaxies and AGN) 
is larger than the assumed critical value 
$J_{\rm cr}=300\times 10^{-21} \, \rm erg/s/Hz/cm^2/sr$ \citep[see][and references therein, 
for a discussion regarding $J_{\rm crit}$]{V17a}. 
As shown in V16, the critical level is reached at $z\sim 16-18$, depending on the 
particular merger history.

We stop forming Pop~III stars and BH seeds at $z<15$, when metal enrichment to $Z>Z_{\rm cr}$ triggers the transition to  Pop~II star formation.

\section{Spectral Energy Distribution models}
\label{sec:SEDmodel}

In this section, we describe how we derive the SED of
galaxies hosting the first seed BHs. 
Once \gamete \, has predicted the physical quantities for each galaxy and its central BH
(i.e. the time-dependent BH mass, BH accretion rate, 
gas density, star formation rate, stellar mass, ISM metallicity and dust-to-gas ratio),
we compute the expected intrinsic emission of
the central accreting BH and stellar component, separately.
Then, we process the combined emission by means of the plasma photo-ionization code
\cloudy, which models the physical conditions of clouds (ionization, chemical and thermal state) 
exposed to emitting sources.
Calculations were performed with version 13.04 of \cloudy\footnote{www.nublado.org} \citep{Ferland13}.

\subsection{Stellar component}

The intrinsic emission from Pop~III stars is computed as the sum of single stellar Black Body (BB) 
spectra, depending on the randomly extracted stellar masses. We compute the BB by using the stellar 
effective temperatures given by \citet{Schaerer02}. 

The SED of Pop II stars are computed with the spectral evolution code 
PEGASE v2.0\footnote{http://www2.iap.fr/users/fioc/PEGASE.html}
(\citeauthor{Fioc97} \citeyear{Fioc97, Fioc99}).
As an input to the code, we provide the Larson IMF (not included in the original version)
and the file containing the star formation and metallicity evolution histories (as a function of time,
for each progenitor galaxy) as predicted by \gamete.

\subsection{Accreting BHs}

The intrinsic SED of accreting BHs has been modelled as described in \citet{P17a}.
The optical/UV primary emission from the accretion disc is the sum 
of multicolor Black Body (BB) spectra given by:
\begin{equation}\label{eq1}
L^{\rm BB}_{\nu} \propto \int^{T_{\rm max}}_{0} B_{\nu}(T) \left( \frac{T}{T_{\rm max}}\right)^{-11/3}  \frac{dT}{T_{\rm max}},
\end{equation}  
where $B_{\nu}(T)$ is the Planck function and $T$ the disc temperature, which reaches the
maximum value, $T_{max}$,
nearby the innermost stable circular orbit (ISCO).
$T(r)$ is the temperature 
profile of steady-state, optically thick, geometrically thin accretion disc described
by \citet[][see \citealt{P17a} for further details]{SS1973}. 
The template spectrum is normalized to the accreting BH bolometric luminosity.

The part of the spectrum in the X-ray energy band due to the emission from the hot corona 
is modeled with a power law with an exponential cut-off at energy 
$E_{\rm c} =300$ keV (\citealt{SOS04, Yue13}): $L_\nu \propto \nu^{-\Gamma+1} e^{-h\nu/E_c}$.
We include the contribution of the metallicity-dependent reflection component 
(the primary X-ray emission reflected by the surrounding neutral medium) using the reflection-only 
solution of the \textsc{pexrav} model \citep*{Magdziarz95} in the \textsc{xspec} 
code\footnote{We consistently linked $\Gamma$, $E_{\rm c}$ 
and normalization parameter to the values adopted for the primary powerlaw component and set
the inclination angle to $60^\circ$ and the reflection strength parameter, $R=1$, 
consistent with typical values of local low-luminosity AGN 
\citep[][Zappacosta et al. 2017; Del Moro et al. 2017]{Burlon11}}.

Following \citet{Brightman2013}, we consistently compute the photon index,
$\Gamma$, in the hard X-ray band ($2-10$ keV) for each accreting BH as a function 
of the Eddington ratio $\lambda_{\rm Edd}$ (predicted by \gamete) as:
$\Gamma_{\rm 2-10 keV} = 0.32 \log\lambda_{\rm Edd} + 2.27$.

\subsection{Emerging spectrum}
The AGN and stellar intrinsic emission components are used 
as an input to \cloudy. The emerging SED is the result of the total intrinsic 
radiation reprocessed by the column of gas, enriched by 
atomic metals and dust, present in the host galaxy.

We assume a static, spherical geometry for the material distribution within the galaxy.
In \cloudy, the spherical gas cloud configuration is described by distributing the material 
in concentric shells with progressively larger radii.
The inner shell face is illuminated by the radiation coming from a combination of point-like sources 
placed at the center.

\begin{figure}
\centering
\hspace{-1cm}
\includegraphics [width=8cm]{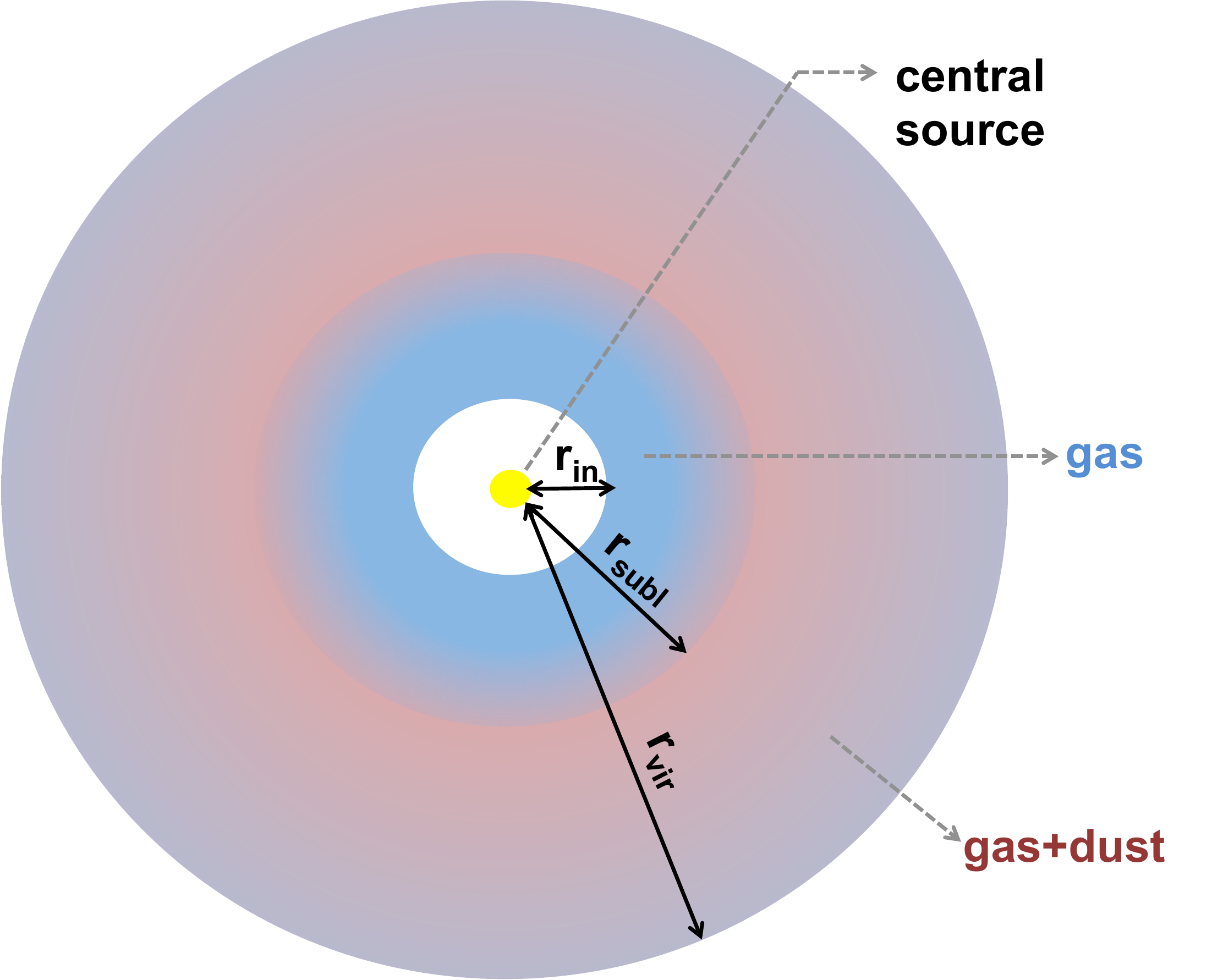}
\caption{Schematic view of the geometry adopted for  \cloudy \, simulations (see text for details).
         Different black solid arrows show the inner, $r_{\rm in}$, sublimation, $r_{\rm subl}$, and DM 
         halo virial radius, $r_{\rm vir}$. Dust free gas is distributed in the inner $r<r_{\rm subl}$ shell,
         while the dusty ISM of the host extends from $r_{\rm subl}$ to $r_{\rm vir}$.}
\label{fig:schema} 
\end{figure} 

To compute the emerging spectrum,  \cloudy \, requires the following set of  
parameters: 
\begin{itemize}
  \item inner and outer radii of the sphere, $r_{\rm in}$ and  $r_{\rm out}$;
  \item gas number density at the inner radius $n(r_{in})$ and the power index, $\alpha$, of the gas radial density profile;
  \item gas composition: column of hydrogen, helium, atomic metals (or gas metallicity) and dust-to-gas ratio.
\end{itemize}
\noindent
In \gamete, the gas density profile is described by a singular isothermal sphere with a flat core: 
\begin{equation}
 n(r) = \frac{n_{\rm norm}}{1+(r/r_{\rm core})^2}.
 \label{eq:gasprofile}
\end{equation}
\noindent
The term $n_{\rm norm}$ is the normalization constant set by the condition that the total 
gas mass of the galaxy is enclosed within the halo virial radius, $R_{\rm vir}$.
A core radius of $r_{\rm core}=0.1 \, (0.01)$ pc has been chosen to encompass the small scale of 
the inner, higher-density regions around the central accreting heavy (light) seed BH\footnote{As a comparison, 
the radius of gravitational influence of a BH, $\sim G M_{\rm BH}/c^2_{s}$ is 
$\sim 3$~pc for $M_{\rm BH}=10^5 \, M_\odot$ ($c_{s} \sim 15 \rm \, km/s$ for $T_{\rm gas}=10^4$~K)
and $\sim 0.03$~pc for $M_{\rm BH}=100 \, M_\odot$ ($c_{s}\sim 5 \rm \, km/s$ for $T_{\rm gas}=10^3$~K).}. 

We adopt the inner radius $r_{\rm in} = r_{\rm core}$ and the number density 
$n(r_{\rm in}) = n_{\rm norm}$
as input values for the \cloudy \, simulations and select the \cloudy \, density profile 
$n(r)=n(r_{\rm in})(r/r_{\rm in})^\alpha \rm cm^{-3}$ with $\alpha=2$, that closely match the profile of
eq.~\ref{eq:gasprofile} adopted in the \gamete \, calculation. 
The outer radius is set to be $r_{\rm out}=R_{\rm vir}$.
A schematic representation of the adopted
geometry is given in Fig.~\ref{fig:schema}.

The host galaxy metallicity and dust-to-gas ratio predicted by \gamete \, are
transferred to the \cloudy \, computation by rescaling the built-in table ``ISM'' 
abundances which accounts for metals and dust mixtures\footnote{In \cloudy \, the ``ISM'' 
gas-phase abundances are from \citet{CowieSongaila86}, \citet{SavageSembach96}, \citet{Snow07}, and \citet{Mullman98} and the dust grain model is from \citet{Mathis77}.}. 

We have verified that dust grains in the very inner regions (i.e. in the shells 
closer to the emitting source) can be heated above the sublimation limit. 
In other words, the maximum temperature of a grain species 
exceeds the default sublimation temperature (1400 K and 1750 K for silicate and graphite grains, 
respectively). Ignoring this aspect may overestimate the IR emission.
To account for dust sublimation, when required, we perform simulations with \cloudy \,
adopting the following three-step strategy (runs):
\begin{enumerate}
  \item We determine the dust sublimation radius, $r_{\rm subl}$ (with $r_{\rm subl}>r_{\rm in}$), at 
        which the temperature of the hottest grain species first falls below the 
        sublimation temperature (run 1);
  \item We process the accreting BH emission traveling through the inner dust-free 
        shell of depth $r_{\rm subl}-r_{\rm in}$. 
        Thus, the inner shell of the cloud is heated only by the AGN (run 2);
  \item We combine the transmitted BH emission, resulting from run 2, with the emission from 
        stellar sources and give the resulting SED as an input for a third \cloudy \, simulation 
        for the outer shells, $r_{\rm subl}\leq r\leq r_{\rm out}$ (run 3).
\end{enumerate}
 
This three-steps simulation strategy allows us to approximately mimic the fact that the AGN and 
stellar emission do not come form the same location in the host galaxy and are expected to be 
reprocessed by different gas and dust column densities. 
It is reasonable to assume that star forming regions are located 
on scales of few up to few hundreds pc \citep[e.g.][]{Davies06,Davies07,Wada09,Hicks09,Sani12} 
where dusty molecular clouds can survive the destructive effects of the strong radiation field of the
central AGN (i.e. in overdense regions where the cold,
molecular gas is shielded by dust).

Finally, we include the redshift-dependent Cosmic Microwave Background radiation field, 
modeled by \cloudy \, as a BB spectrum with temperature $T_{\rm CMB}=T_0 (1+z)$ K 
\citep{Peebles71, Mather99, Wilkinson87}.

\begin{figure}
\centering
\includegraphics [width=8.5cm]{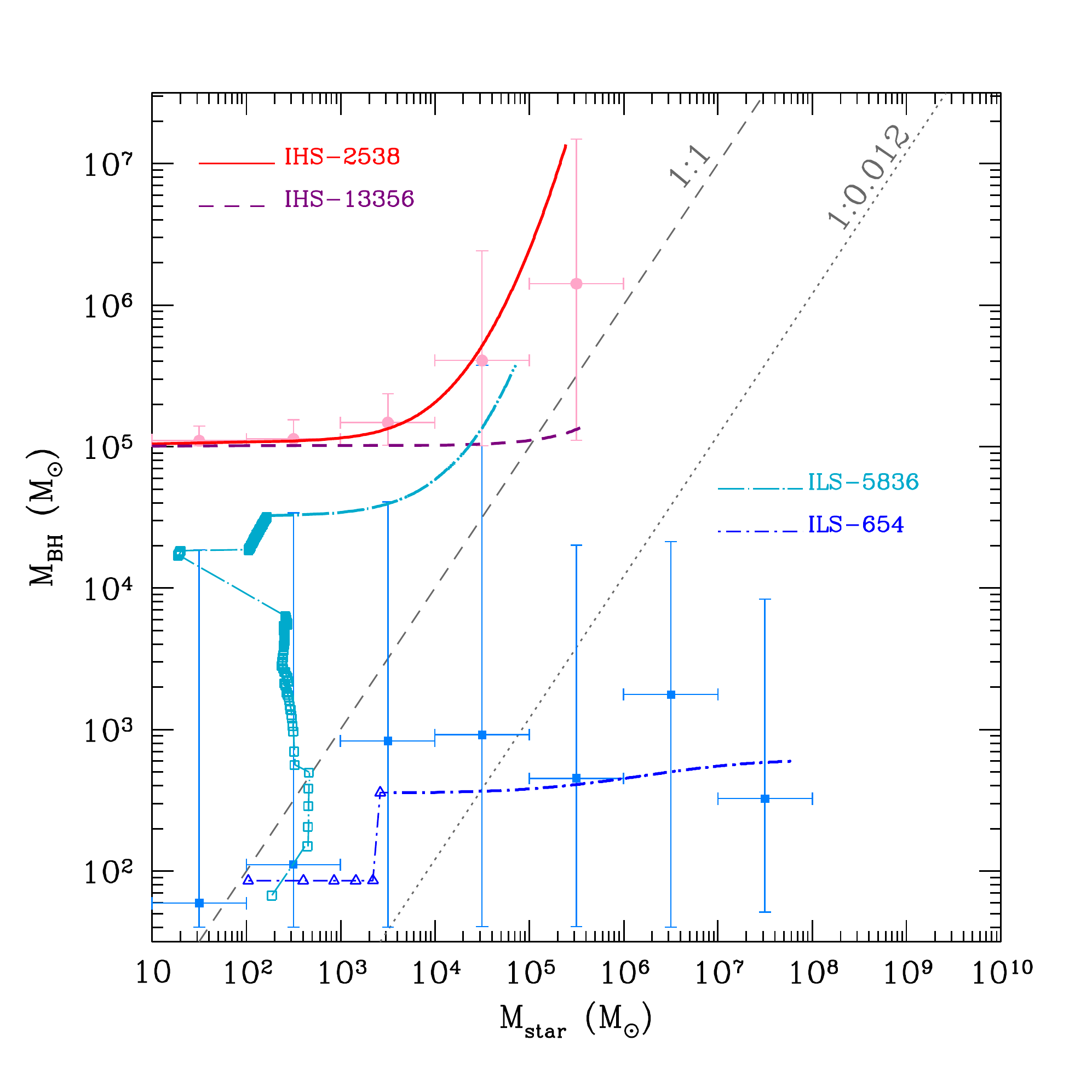}
\caption{The BH mass as a function of the stellar mass for the selected accreting seed BHs prototypes.
Solid and dashed lines are IHS-2538 and 13356, respectively. 
Two ILS are instead shown with dot-dashed (654) and dot-long-dashed (5836) curves. For these systems,
data points indicated by squares (triangles) indicate the evolution during the Pop~III star 
formation regime.
We also label the fraction of systems in the IHS and ILS populations showing properties similar 
to each prototype (see text for details).
Filled light blue squares and pink circles represent, respectively, 
the average BH mass of ILS and IHS in different stellar mass bins,
with bars indicating the maximum and minimum BH mass in each bin and the width of the stellar mass 
interval. The dotted and short-dashed lines indicate the 1:1 and local BH-stellar mass
\citep{Sani11} scaling relations.}
\label{fig:selectedBHs} 
\end{figure} 

\section{RESULTS}

In this section, we compute the time-dependent SED
of four proto-typical ILS and IHS along their evolution. Since our main aim is
to investigate if the nature of BH seeds can be observationally discriminated,
we select two efficiently-growing and two inefficiently-growing BH seeds among 
the IHS and ILS samples. In this way, we highlight differences
in the same population of systems and similarities among the two different classes.
In what follows, we indicate each system 
by  its host halo ID assigned in the simulation merger tree.

\subsection{Most and least massive BH in the sample}

{
We have selected the most and least massive BH, in the sample analyzed in paperI.
The physical and observational properties of these systems are representative of sources 
that we expect to find in biased regions that lead to the formation of a SMBH at $z>6$.
}

Fig.~\ref{fig:selectedBHs} shows the evolution of selected ILS and IHS systems 
in the BH-stellar mass plane.
The differences among the four tracks are indicative of the large variety of 
evolutionary channels that we find in the simulation. 

The two IHS (2538 and 13356) are shown by the solid and
dashed lines. Initially, they follow a similar evolution at constant BH mass. When the
stellar mass of the hosts has reached a value of $~ 10^3 \, M_\odot$, 
the BH mass of 2538 grows faster than the stellar mass, reaching a value of $\sim 10^7 \, M_\odot$,
whereas 13356 continues to evolve at an almost constant BH mass. It is interesting to
compare the evolution of these two systems with the average evolution of IHS, shown
by the pink circles data points. It shows that 2538 better represents the average 
evolution of IHS hosted in galaxies with $M_{\rm star} \leq 10^{4.5} \, M_\odot$. 
The fraction of systems similar to IHS-2538 ($\sim 33\%$) and 13356 ($\sim 40\%$) are also
labelled in the figure. The remaining fraction of objects instead shows mixed properties
in terms of final BH mass and/or lifetime in isolation ($\Delta t$, see Figure 2 in paperI).

The two ILS (5836 and 654) are shown by dot-dashed and dot-long dashed 
lines, respectively. Triangle-shaped (squared) data points indicate their growth during 
the Pop~III star formation epoch (see Appendix~\ref{appendix:prop} for further details).
They start with comparable BH mass but follow a very different evolution. 
ILS-5836 grows much faster than the stellar mass (squares) until Pop~II stars start to form in 
the host halo (dot-long dashed line) whereas ILS-654 evolves at an almost constant BH mass.
In both systems the central BH experience mergers with other Pop III BHs formed in the same host 
galaxies (see Section 3.2 in paperI).
These events cause the step-like increase in the BH mass along the two tracks 
(see Appendix~\ref{appendix:prop}).
The comparison with the average evolution of ILS, shown by the light blue squares, 
indicates that ILS-654 is more representative of the average ILS evolution.
{About $89\%$ of systems are similar to ILS-654 while systems following an evolution 
(in terms of BH mass) similar to ILS-5836 are only $\sim 1\%$ of the 
population\footnote{Here we are considering all 
systems with $M_{\rm BH}>2\times 10^4 \, M_\odot$ for this estimate}, as labelled in the figure.}

{
A detailed description of evolutionary histories of the four prototypical systems is given 
in Appendix~\ref{appendix:prop}.

We define the age of each system as the time it spends evolving
in isolation from the formation of the BH seed ($\rm age=0$). 
The isolated evolution ends when the first (minor or major) 
halo merger occurs.
It is worth mentioning that although we follow systems in the ``isolated'' phase, this 
does not imply that the selected host halos evolve in isolation prior to seed BH formation. 
Thus, the initial conditions (e.g. gas mass and metallicity at $\rm age=0$) 
are different in different systems, even in those belonging to the same class. 
This reflects the different histories prior to BH seed formation and the host halo selection:
ILS are initially hosted in star-forming mini halos, while IHS hosts are $\rm Ly\alpha-$cooling halos.\\

The four selected systems account for the large diversity in the properties of ILS and IHS 
populations. By looking at individual evolutionary histories (see Figure~\ref{fig:prop}), we infer that BH growth
is affected by \textit{(i)} the environmental conditions at the epoch of BH seed formation 
(i.e. the properties of the host halo resulting from its earlier evolution) and, afterwards, 
\textit{(ii)} by the host galaxy star formation and chemical enrichment histories. 
The final BH mass, at the end of the isolated evolution, is largely determined by these 
properties. 
Systems with similar (ILS-5836 and 654) or even equal (IHS-2538 and 13356) initial seed mass 
can follow very different evolutionary tracks 
reaching final BH masses that differ by $\sim 2$ orders of magnitude.
Systems hosting BHs of different origins (IHS-2538 and ILS-5836) may instead end-up having similar properties
(during certain stages of their evolution).
These differences/similarities are reflected in the predicted SEDs.
}

\subsection{Emission from BH seeds and their hosts}

\begin{figure*}
\centering
\includegraphics [width=8.5cm]{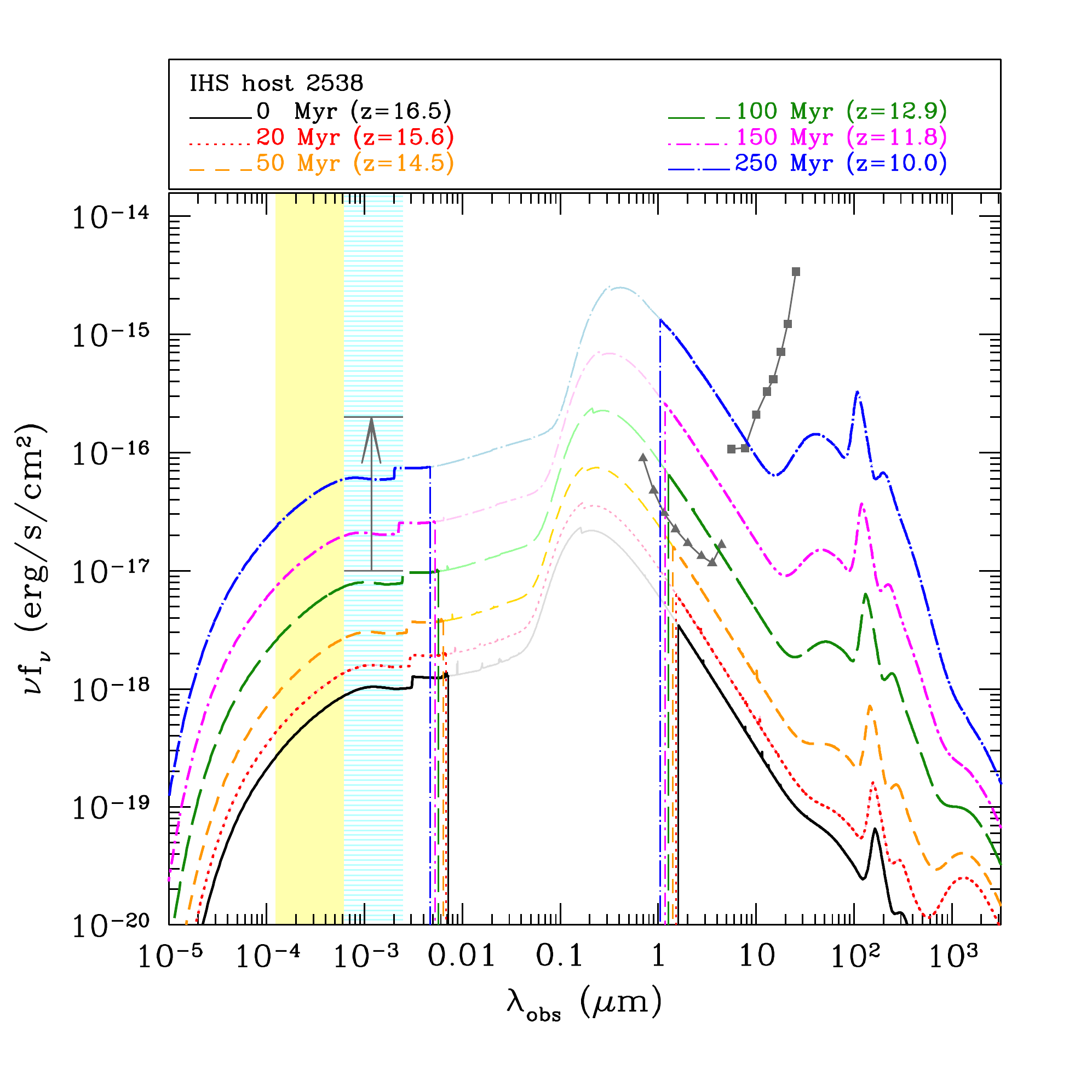}
\includegraphics [width=8.5cm]{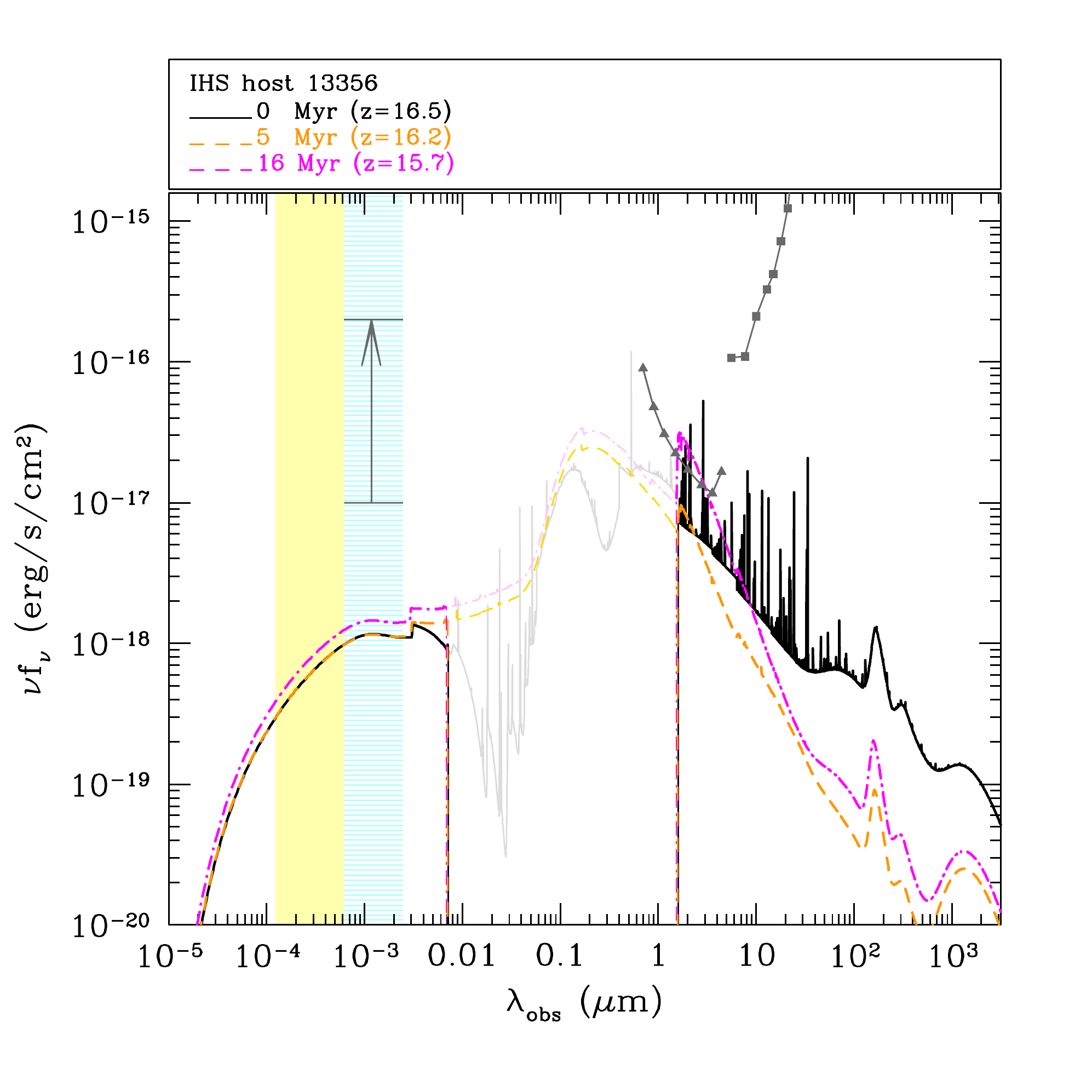}
\includegraphics [width=8.5cm]{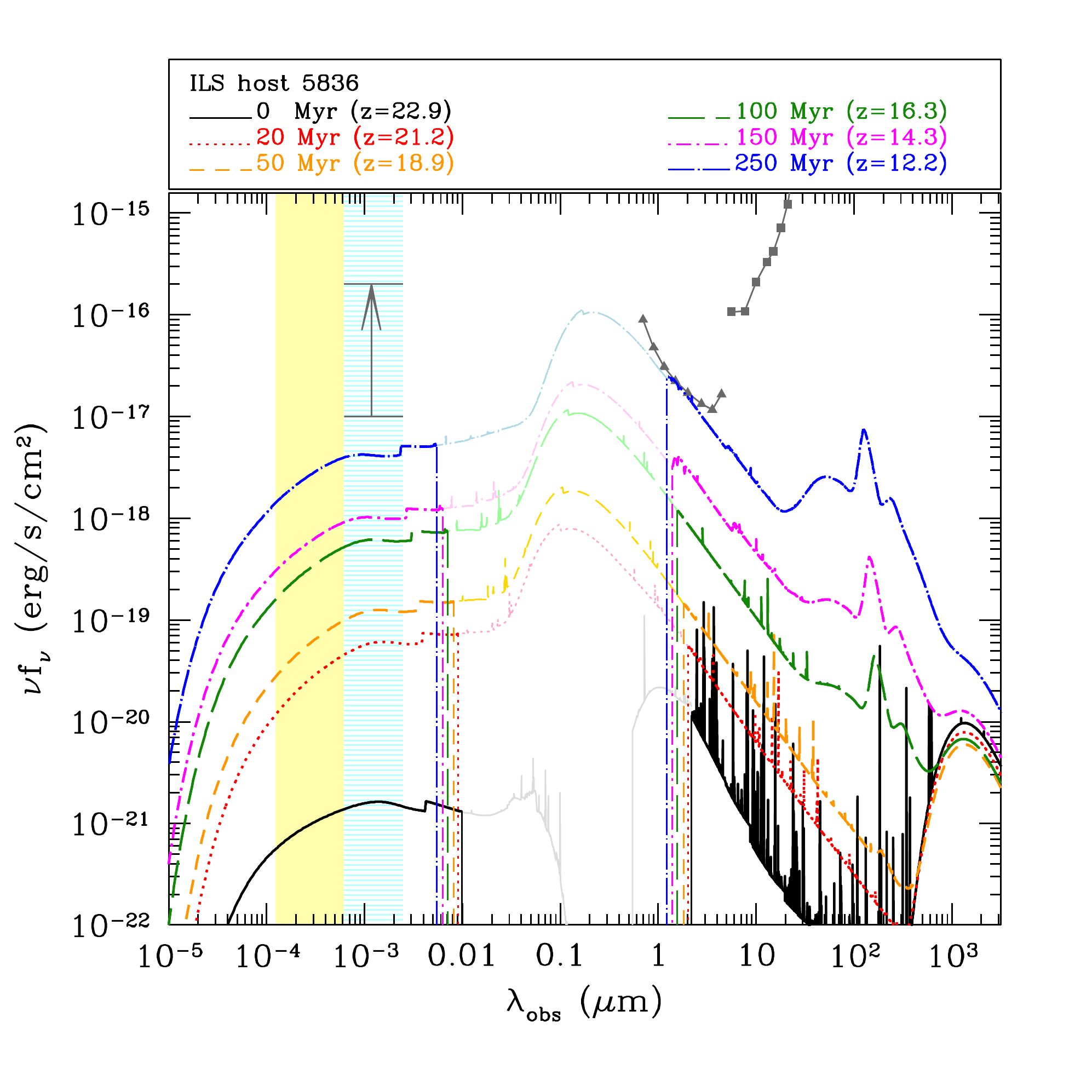}
\includegraphics [width=8.5cm]{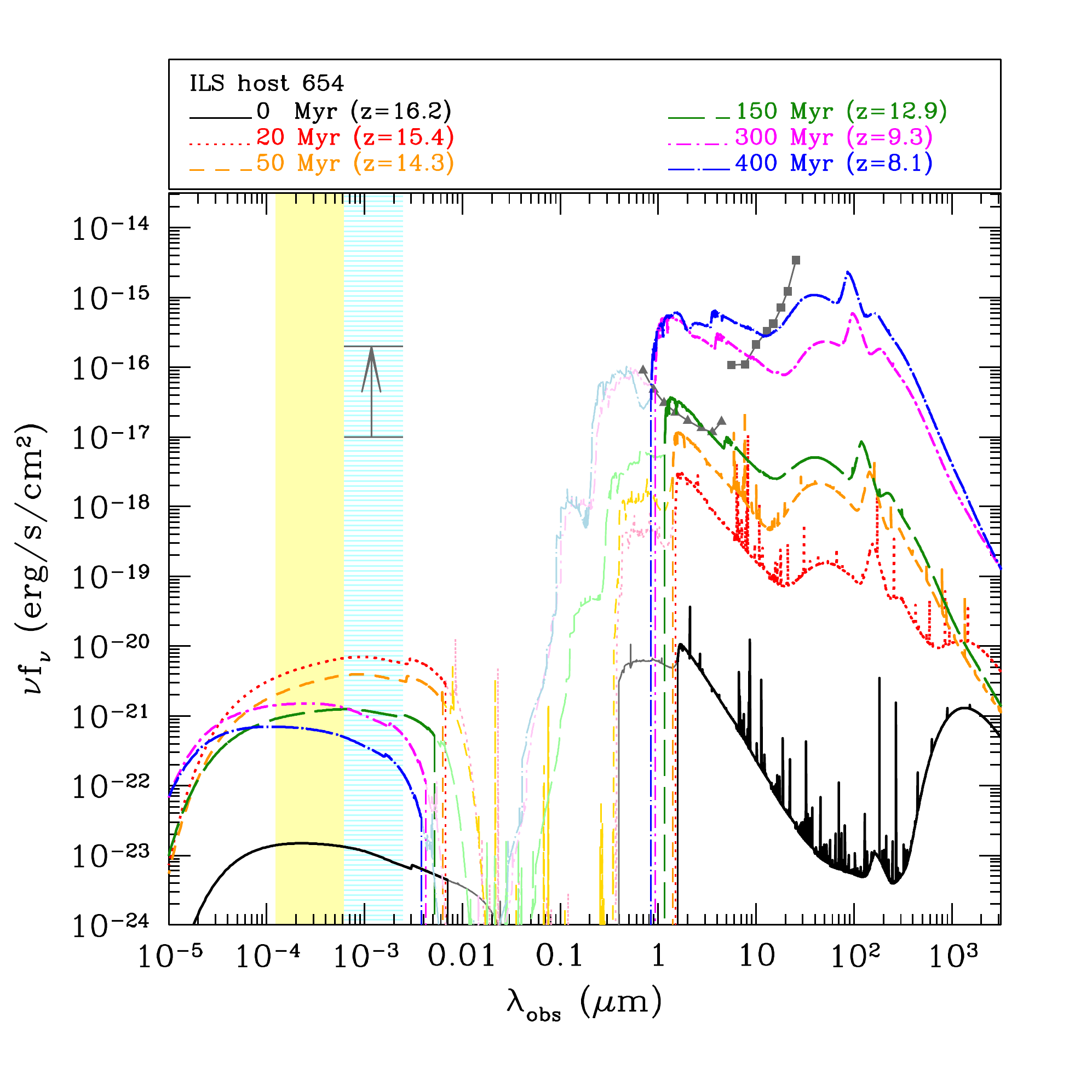}
\caption{ 
SED of IHS-2538 (top-left), IHS-13356 (top-right), ILS-5836 (bottom-left) and ILS-654 (bottom-right)
as a function of the observed wavelength, $\lambda_{\rm obs}$. In each panel, different lines corresponds
to different ages of the system, as shown by the legenda, where we also label the corresponding redshift in the simulation.
The SED are shown by lighter colours at wavelengths where the emission will be completely absorbed by the intervening
neutral hydrogen along the line-of-sight.
The yellow shaded and cyan dashed regions indicate the observed hard and soft X-ray bands, respectively.
The vertical arrow marks the flux limit of future surveys designed for the ATHENA mission by \citet{Aird13}.
The gray triangles and squares are the 
photometric flux limits of the instruments NIRcam and MIRI on the JWST.} 
\label{fig:sed} 
\end{figure*}

Following the procedure described in Section \ref{sec:SEDmodel}, we compute the SED of the selected IHS and ILS presented above 
at different times of their evolution (ages). Clearly, these correspond to different redshifts in the simulation. The results are shown in 
Fig.~\ref{fig:sed}. The SED are represented by lighter colours 
at wavelengths where the emission will be completely absorbed by the intervening neutral hydrogen along the line of sight. These
correspond to rest-frame wavelenghts shortward of the Lyman limit at $912~\AA$, down to about $\sim 100~\AA$ \citep[][and references therein]{Madau95,Inoue14}.
In each panel, the yellow shaded and cyan dashed regions indicate the observed hard, [2-10]~keV, and soft, [0.5-2]~keV, 
X-ray bands. 
The gray arrow in the X-ray part of the spectrum identifies the range in the limiting fluxes
of the ATHENA survey designed by \citet{Aird13} for a  $\sim 1 \, \rm deg^2$ (lower end) and 
$\sim 100 \, \rm deg^{2}$ (upper end) area\footnote{This survey has been designed as a Wide Field Imager 
wedding cake strategy with single tiers of: $\rm 4\times 1Ms$, $\rm 20\times 300 ks$, $\rm 75\times 100 ks$ 
and $\rm 259 \times 30 ks$. Reported flux limits are for a $3^{\prime\prime}$ PSF. A collecting area of $2 m^2$
at 1 keV and an instrument field of view of $40'\times 40'$ have been assumed.}.
The gray triangles and squares report the NIRcam and MIRI photometric 
sensitivities on board of JWST\footnote{These are the photometric performance requirements of the instruments, namely, the point source faintest fluxes that can be obtained at a signal-to-noise ratio SNR =10 in a 10ks 
integration (https://jwst.stsci.edu/science-planning/proposal-planning-toolbox/sensitivity-overview).}.

\subsubsection{The most massive systems}
We first analyze the time-dependent spectra of IHS-2538 (top-left) and ILS-5836 (bottom-left), which 
have been selected to be the most massive BHs among the IHS and ILS samples (see Fig.~\ref{fig:selectedBHs}). 
At each epoch, the spectrum of IHS-2538 is dominated by the AGN continuum emission, resulting in an 
almost featureless SED (no prominent, luminous emission lines can be identified). 
Although strongly attenuated by the intervening material, the most prominent feature in the rest-frame SED 
is the optical/UV emission from the BH accretion disc (the Big Blue Bump, BBB). The reprocessed AGN emission 
extends to $\lambda_{\rm obs} \sim 10 \, \rm \mu m$\footnote{We do not 
include polycyclic aromatic hydrocarbon (PAHs) particles in the computation, which are
expected to radiate in the $[1-13]~\mu m$ range.}.
At longer wavelengths, the SED is dominated by dust thermal emission.  
At $\rm age \geq 100~Myr$, when the BH mass is $M_{\rm BH} > 7 \times 10^5 \, M_\odot$
and the BH accretion rate (BHAR) is $\dot{M}_{\rm BH}> 0.01 \, M_\odot/{\rm yr}$ (see the top-left panel in Fig.~\ref{fig:prop}), 
the rest-frame optical-UV emission is redshifted to the near IR (NIR) bands and is detectable 
by NIRcam.  At almost the same time the emission in the X-ray bands becomes intense enough to 
exceed the lower flux limit in the ATHENA survey.
 
The duration of the isolated evolution of ILS-5836 is comparable to that of IHS-2538 ($\sim 225$~Myr), but this system is
located at a much higher redshift. It forms at $z \sim 22.9$ and evolves in isolation 
down to $z \sim 12$. At $\rm age = 0$, when the first Pop~III stars form, the emission from
the accreting $\sim 100 \, M_\odot$ BH is comparable to that coming from the stellar component.
As the system evolves in time, the emission properties appear to be a scaled-down version of 
IHS-2538 (see also the bottom left panel of Fig.~\ref{fig:prop}).
However, more prominent emission lines are present. 
This is a consequence of the larger gas column density in the host galaxy of ILS-5836, due to the less efficient BH feedback. At $\rm age = 250$~Myr, although its flux is still below the JWST limit, the SED of ILS-5836 is potentially 
indistinguishable, in terms of shape and spectral features, from that of IHS-2538.

\subsubsection{The least massive systems}
Quite different spectra are predicted for IHS-13356 and ILS-654, 
 which have been selected to host the two inefficienly-growing BHs among the IHS and ILS sample
(see Fig.~\ref{fig:selectedBHs}). 
The SEDs of IHS-13356 are shown in the top-right panel of Fig.~\ref{fig:sed} at $\rm age=0, 5$ and 16 Myr.
At $\rm age = 0$, the only emitting source is the newly planted BH. Its radiation travels through a 
large gas column density, as the gas mass is $\sim 20$ times higher than in IHS-2538 at the same epoch
(as discussed in the previous section). 
Among the several emission lines, H and He recombination lines are present. 
Although the continuum level is below the NIRcam photometric capability, some of the lines
are intense enough to be detected by NIRspec, despite the very high redshift of the source. 
As the gas mass drops, at $\rm age \geq 5~Myr$, the spectra are dominated by the continuum.
At 16 Myr, the part of the spectrum (magenta dot-dashed line) entering in the NIRcam filters at 
$1 \, \mu {\rm m} < \lambda_{\rm obs} <3 \, \mu {\rm m}$ is dominated by the stellar component 
($M_\ast \sim 10^5 \, M_\odot$) and it is just above the sensitivity limits. 
At this stage,IHS-13356 emission could be revealed by NIRcam (F200W and F277W filters) 
onboard of JWST.
This low-mass, inefficiently accreting BH, is predicted to be too faint to be detected in the 
X-ray bands, even with ATHENA sensitivity.

Finally, the bottom right panel of Fig.~\ref{fig:sed} shows the SED of the starburst dominated ILS-654.
Similarly to IHS-13356, this is a slowly accreting light BH seed whose mass never exceeds $\sim 10^3 \, M_\odot$,
even after 400 Myr of evolution. The continuum emission is dominated by the stellar component
with the low-mass AGN contributing to the X-ray part of the spectrum, but the emission 
remains several orders of magnitude fainter than any current (and future) instrument sensitivity.
As the galaxy evolves ($\rm age > 100~Myr$) a continuum-dominated spectrum emerges. 
Such an intense stellar emission may be easily revealed by the JWST. 

\subsubsection{Detectability}
To summarize, systems like IHS-2538 or ILS-654 in the latest stages of their evolution 
($\rm age\gtrsim 100$ Myr) could be easily detected by the JWST instruments NIRcam 
(at $\lambda\geq 0.9~\mu m$) and MIRI (at least in the two higher-sensitivity photometric filters,
F560W and F770W). The NIR and mid-IR (MIR) luminosities of these objects are
$L_{\rm NIR}\gtrsim 10^{42} \, \rm erg/s$ and $L_{\rm MIR}\gtrsim 10^{43} \, \rm erg/s$, 
respectively\footnote{These are the NIR and MIR band luminosities provided by \cloudy 
in the wavelength ranges $[1-5] \, \rm \mu m$ and $[5-40] \, \rm \mu m$, respectively.}. 
On the other hand, detecting objects like ILS-5836 
would instead require deeper NIRcam campaigns. 
Interestingly, we find that the emission from a pure accreting heavy seed (no stars) enshrouded in a dense gas 
cloud, as in IHS-13356, may be revealed already at $z>16$ by the spectrograph NIRspec, thanks 
to its prominent emission lines. The stellar-dominated radiation emerging afterwards (at $z\sim 15.7)$
may be instead detected in the F200W and F227W NIRcam filters.
Finally the ATHENA survey program envisioned in \citet{Aird13} will reach flux limits which are
exceeded only by accreting BH seeds similar to IHS-2538 at $age> 100$ Myr, when their soft X-ray luminosity 
becomes $L_{[0.5-2] \rm keV}\geq 10^{42} \, \rm erg/s$.

We conclude that light and heavy seeds and their hosts can have very different emission properties. 
However, these appear to be more significantly affected by their individual evolutionary history rather 
than by their initial nature, i.e. whether they classify as a light or heavy BH seed. 
Yet, we recall that the two ``antithetical'' seed proto-types, 
IHS-2538 and ILS-654, show very different emission spectra: the first one being dominated by the  
central AGN and the second one being dominated by the stellar component of the host galaxy.
While ILS-654 better represents the average evolution of light seeds, 
IHS-2538 represents $\sim 33\%$ of the IHS population. Still, systems like IHS-2538 are 
the most luminous among the IHS, thus, having the largest probability of being detected by JWST.

\subsection{The mean IHS}
{
As discussed above, the four selected prototypes represent extreme cases (the most and least 
efficiently growing IHS and ILS). In the ILS population, the results obtained for ILS-654 are also 
representative of the average population (see Figure~\ref{fig:selectedBHs}). 
Thus, in order to make our predictions for IHS systems more general, we perform an additional 
analysis, by investigating the SEDs and observational properties of the ``average'' IHS population. 
To this aim we have constructed and processed AGN and stellar emission obtained by considering the 
average IHS evolution shown in Figure~\ref{fig:selectedBHs} (pink data points). 
The evolution of the mean IHS host has been weighted over the average lifetime in isolation, 
$\sim 65$ Myr (obtained from the distribution discussed in paperI)\footnote{For the mean IHS, age$=0$,
corresponds to the average formation redshift of heavy seeds, that is $z\sim 16.5$, the same redshift at 
which IHS-2538 forms. The final redshift, after $\sim 65$ Myr of isolated evolution, corresponds 
to $z\sim 13.9.$}. 

The mean IHS is an AGN dominated system and its evolution is very similar (in both normalization and spectral features) to IHS-2538,
although its final BH is less massive ($\sim 1.4\times 10^6 \, M_\odot$ at $65$ Myr). 
For this reason, we do not show the predicted time-dependent SED. 
In particular, the final stage of the mean IHS (i.e. at $age=65$ Myr) its SED is comparable to 
that of IHS-2538 at $age=100$ Myr (with the mean IHS, being potentially detectectable by JWST). 
This is consistent with the fact that at these stages, when the BH mass is $\lesssim 10^6 \, M_\odot$, 
IHS-2538 and the mean IHS grows following similar evolutionary tracks.
}

\begin{figure}
\centering
\includegraphics [width=8.5cm]{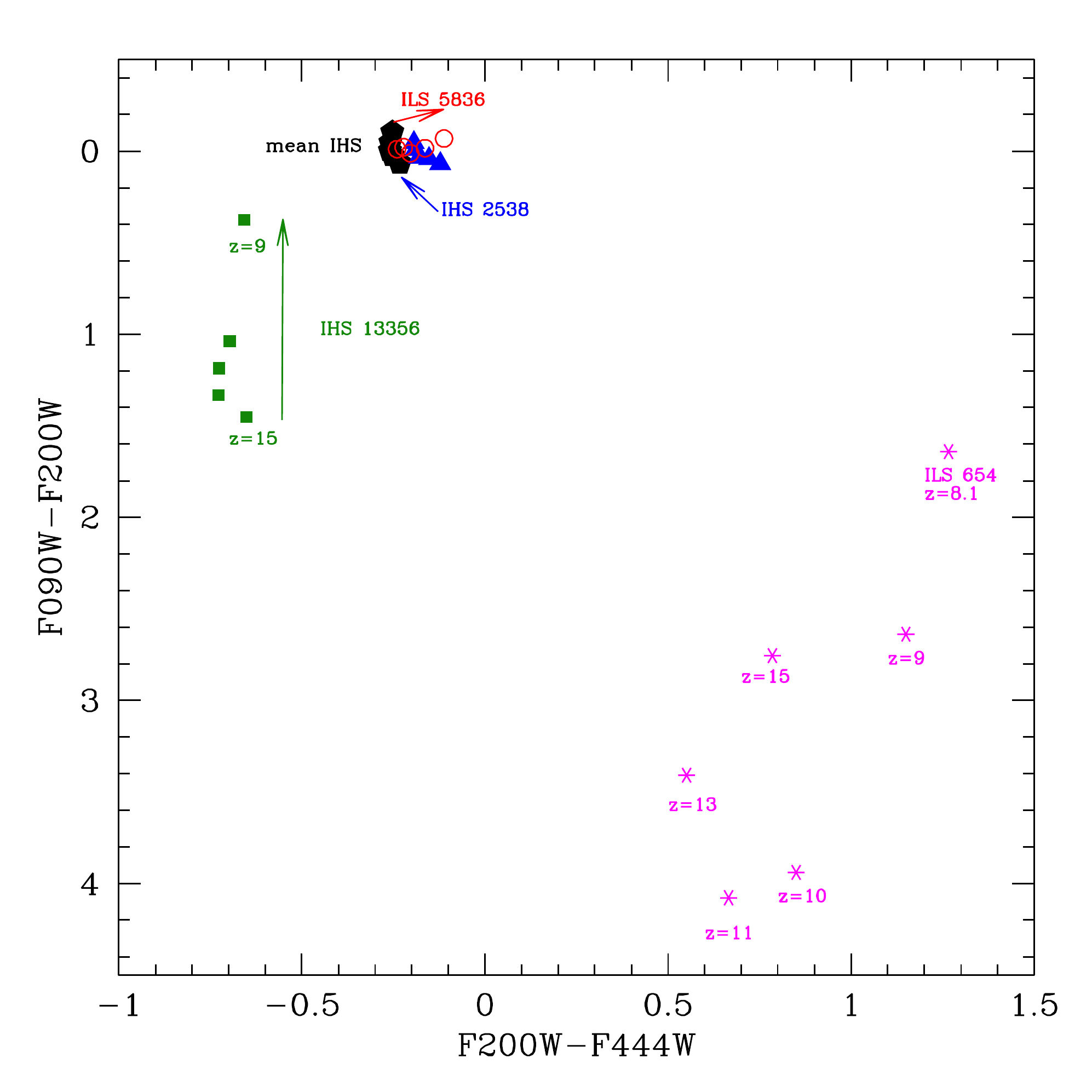}
\caption{The color-color diagram for the selected proto-types at 
$z = 9, 10, 11, 13$ and $15$. IHS like 2538 and 13356 are shown by triangles and squares, respectively.
{Black pentagons are for the mean IHS case}.
ILS like 5836 and 654 are shown by circles and asterisks, respectively.
For ILS-654 we also show the case $z=8.1$, corresponding to the redshift 
at which the host reaches its maximum luminosity, at $\rm age=400$ Myr 
(see text). For other sources the arrows indicate the decreasing redshift direction.} 
\label{fig:colors} 
\end{figure} 

\subsection{Observational features of BH seeds and their hosts}

To investigate the observability of growing BH seeds, we consider the 
systems presented above, compute their colours in the NIRcam
filters and their X-ray luminosities.

\begin{table}
\caption{Bolometric luminosities of systems during the brightest stage of their evolution (see text for details).}
\centering
\begin{tabular}{c c c c}
\hline
  {\bf name} & {\bf age (Myr)} & {\bf redshift} & {\bf $L_{\rm bol} \, \rm erg/s$}\\
\hline
  IHS-2538  & 250  & $\sim 10$   & $\sim 10^{45}$ \\
  IHS-13356 &  16  & $\sim 16$   & $\sim 4\times 10^{43}$\\
  mean IHS  &  65  & $\sim 13.9$ & $\sim 9.2\times 10^{43}$\\
  ILS-5836  & 225  & $\sim 12$   & $\sim 5\times 10^{43}$\\
  ILS-654   & 405  & $\sim 8.1$  & $\sim 3.5\times 10^{44}$\\
\hline
\end{tabular}
\label{tab:lbol}
\end{table}

To this aim, we first take the SED of each system during
its brightest phase, which corresponds to different ages and redshifts 
(see Fig.~\ref{fig:sed}). {Bolometric luminosities of systems at the
brigthest stage are given in Table~\ref{tab:lbol}}.
%
During these phases, all the systems reach AB magnitudes $M_{\rm AB}<29$, 
above the JWST sensitivity limit for a 10 ks exposure in the $[0.7-4.4]~\mu$m spectral 
band\footnote{The four selected systems are F090W drop-outs, as the Lyman limit ($912 \, \AA$) 
is redshifted to longer wavelengths, with the exception of 
ILS-654 at $\rm age\geq 300$ Myr (i.e. at $z\leq 9$).}.

Then, we shift the above SED at $z=9, 10, 11, 13$ and $15$, in order to 
compare the four systems at the same redshift\footnote{At $z\geq 10$ these must be 
considered as upper limits as the Lyman-limit ($912\,\AA$ rest frame) falls in the F090W 
filter at these redshifts.}.
This approach enables us to provide more general predictions, independently of the redshift
at which the brightest stage is reached.

{To compare with previous studies, we adopt the same colour-colour diagram 
proposed by \citet{Natarajan17}}.
The resulting F090W-F200W - F200W-F444W color diagram is shown in Fig.~\ref{fig:colors}.
The predicted colors reflect different galaxy types, 
with AGN-dominated systems (such as IHS-2538, {the mean IHS} and ILS-5836) being bluer than starbursts 
(ILS-654). 
The two classes are clearly distinguishable in the color-color plane, independently of 
their redshifts. Despite its lower AB magnitude\footnote{ILS-5836 reaches 
$\rm M_{AB}\sim 29$ only at the end of its isolated evolution, at $z\sim 12$, and 
would be detectable with NIRcam at lower redshifts.} 
we predict ILS-5836 to have 
very similar colors to IHS-2538. 
Hence, independently of their nature and of their redshifts, we find that efficiently
growing BH seeds occupy a well defined region of the color-diagram. 

\begin{figure}
\centering
\includegraphics [width=8cm]{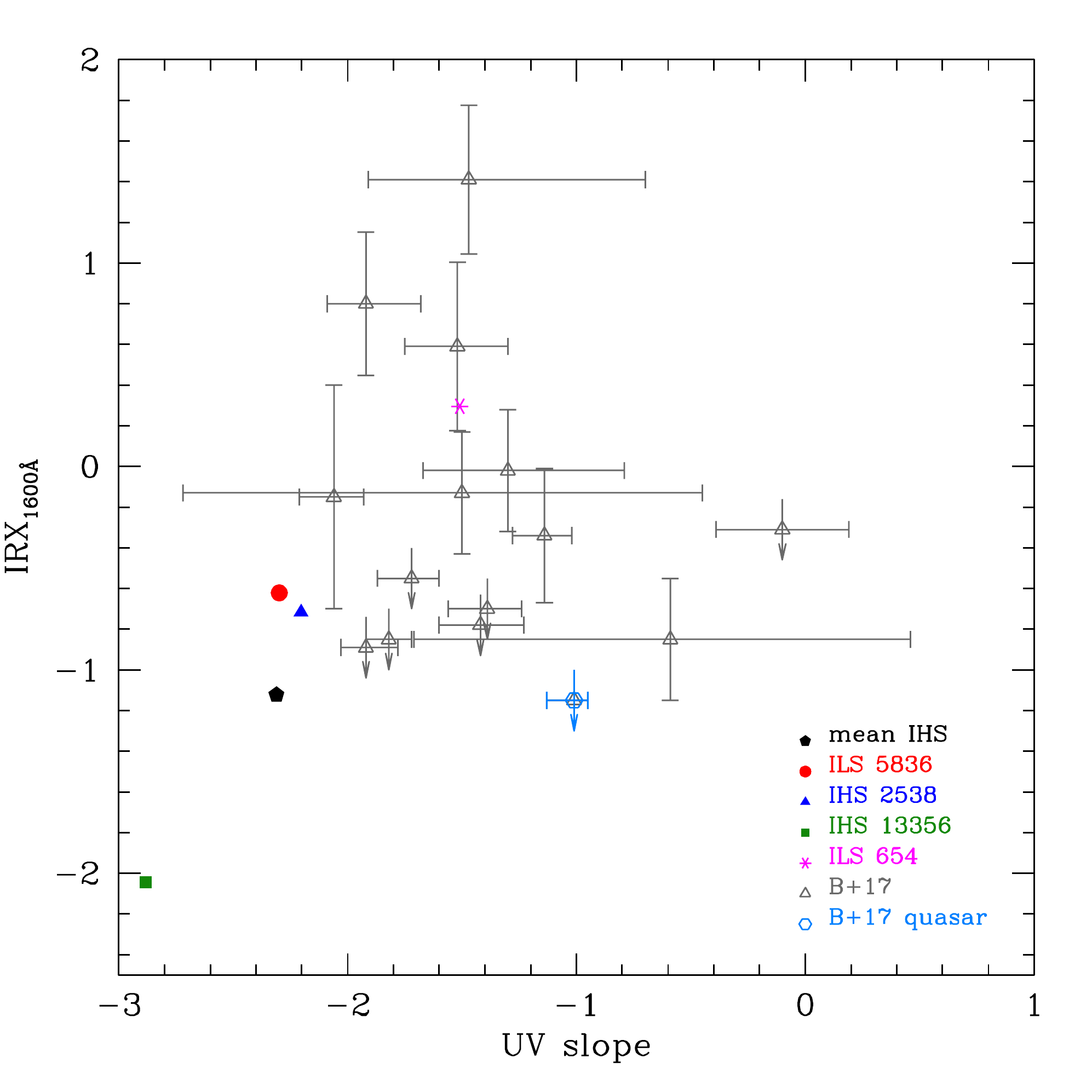}
\caption{The IRX as a function of the UV continuum slope. 
Filled symbols represent IHS (2538 blue triangle, 13356 green square, {mean IHS black pentagon}) and
ILS hosts (5836 red circle, 654 magenta asterisk). 
Empty data points with error bars are from \citet{Barisic17}, with the hexagon 
indicating a quasar candidate at $z=5.3$.} 
\label{fig:irx} 
\end{figure} 

In Fig.~\ref{fig:irx} we show the IRX--UV slope diagram for the same systems and compare
the results with the sample of $z\sim 5.5$ galaxies presented by \citet{Barisic17}.
We compute the ratio between the rest frame IR and UV luminosities as
IRX = log $(L_{\rm IR}/L_{1600\AA})$ and the  UV continuum slopes, $\beta$, 
as a least-square linear fit in the $[1500-3000] \, \AA$ part of the spectra.
Since  the IRX is related to dust extinction, it is not surprising that
the dustiest ($D_{\rm ISM}\sim 5\times 10^{-4}$, see Fig.\ref{fig:prop}) 
starburst prototype ILS-654 shows a high IRX, $0.29$, and a spectral slope 
$\beta=-1.51$, consistent with some of the $z\sim 5.5$ observed galaxies 
\citep{Mancini16, Barisic17}. 
The IR luminosity (integrated from $8~\mu$m to $1000~\mu$m) of this system is 
$L_{\rm IR}\sim 10^{44} \, \rm erg/s$ ($\sim 85\%$ of the bolometric luminosity).

As expected, based on the similarity of their SED and colors, IHS-2538 and ILS-5836 
have similar IRX and UV slopes. In these systems,
the UV light from the accreting BHs is only mildly absorbed by the poorly enriched ISM
of their host galaxies (see Fig.~\ref{fig:prop}), resulting in a lower IRX, $\sim 0.6-0.7$, 
than in ILS-654. {The mean IHS show a lower IRX with respect to IHS-2538 
as a consequence of the lower associated dust-to-gas ratio (estimated by averaging 
over the dust-to-gas ratio of all the IHS systems). 
With respect to ILS-5836, the IRX of mean IHS has is lower because of its higher $L_{1600\AA}$.
}

Finally, the exceptionally blue slope ($\beta \sim -2.88$) and low IRX ($\sim -2.04$) 
of IHS-13356 reflect the almost unattenuated, pure stellar emission
dominating the SED at its maximum. The predicted $\beta$ slope is consistent 
with the mean color of faint $z \sim 7$ galaxies measured by \citet{Bouwens10}.

\begin{figure}
\centering
\includegraphics [width=9cm]{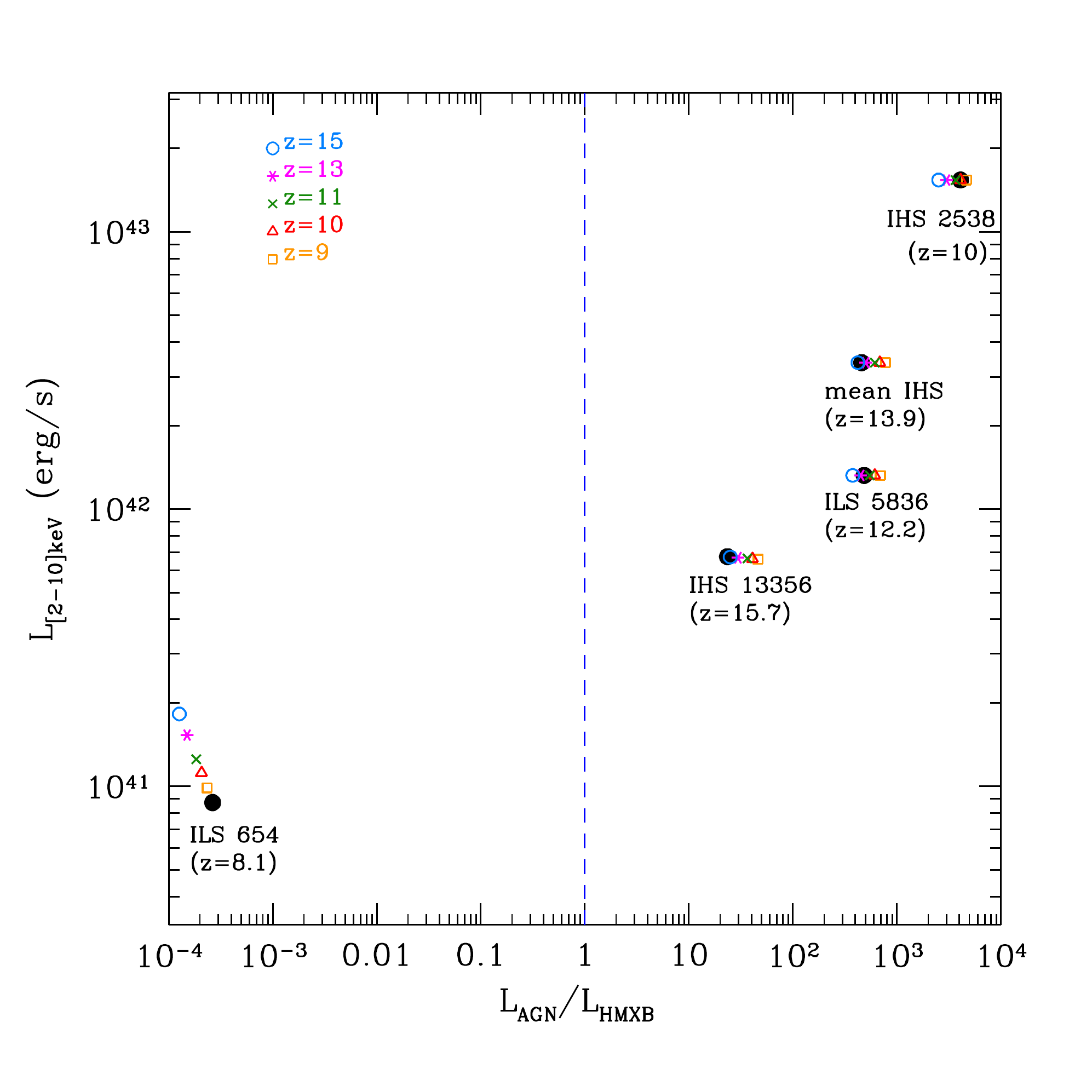}
\caption{The relative contribution of accreting BHs and HMXBs to the luminosity at $[2-10] \rm keV$
for the four prototypes in our sample: IHS-2538, IHS-13356, ILS-5836 and ILS-654, {and the mean IHS}. 
Black filled circles show the luminosity at the original redshift of the system while empty coloured 
symbols indicate the predictions at $z=9$ (yellow squares), $10$ (red triangles),
$11$ (green crosses), $13$ (magenta asterisks) and $15$ (azure circles). 
The blue vertical line depicts an equal contribution from AGN and HMXBs.} 
\label{fig:hmxb} 
\end{figure} 
The properties discussed above can be used to identify accreting BH seed candidates 
in future JWST surveys, distinguishing them from starbursts.

By combining the information obtained from color-color and IRX-$\beta$ diagrams
we suggest that the best candidates for 
future and deeper X-ray follow-up observational campaigns
should be selected according to the following conditions:

\begin{itemize}
  \item $-0.3\lesssim$F200W-F444W$\lesssim 0$ and F090W-F200W$\gtrsim -0.2$ 
  \item UV continuum slopes $\beta<-2$ and $-1.5<$IRX$<0$.
\end{itemize} 

One additional limitation to the observational identification of growing BH seeds is that 
the X-ray luminosities of faint BHs is hard to discriminate from the cumulative emission of 
high-mass X-ray binaries (HMXBs) in the host galaxies  \citep[e.g.][]{Volonteri17}. 
In Fig.~\ref{fig:hmxb} we show the hard X-ray band luminosity,
$L_{\rm [2-10]keV}=L_{\rm AGN}+L_{\rm HMXB}$, as a function of the ratio between the AGN and HMXBs 
luminosities, $L_{\rm AGN}/L_{\rm HMXB}$, in the same spectral band. 
Following \citet{Volonteri17}, $L_{\rm HMXB}$ is computed using the empirical relation 
found by \citet{Lehmer16},
log$(L_{\rm HMXB} = 39.82 + 0.63$log$\dot{M}_{\ast} + 1.31$log$(1+z)$, at the original redshift 
of the system in the simulation and at redshifts $z = 9, 10, 11, 13$ and $15$\footnote{The contribution of HMXBs has been computed only for the purpose of this comparison and it is not 
included in the SED models presented in Fig.~\ref{fig:sed}.}.

We find that systems with properties similar to IHS-2538, {the mean IHS} and ILS-5836, 
have $L_{\rm AGN}/L_{\rm HMXB}>100$. These would be easily identified as AGN.
Although the SED of IHS-13356 is dominated by stellar emission (see Fig.~\ref{fig:sed}) in the optical/UV 
(rest frame) part of the spectrum, a star formation rate of $\dot{M}_\ast \sim 0.027 \, M_\odot/$yr 
(see Fig.~\ref{fig:prop}) corresponds to population of HMXBs that produce less than 
$\sim 5\%$ of the total luminosity at $[2-10]$ keV ($L_{\rm AGN}\sim 6.5\times 10^{41} \, \rm erg/s$ 
while $L_{\rm HMXB} \sim 2.7 \times 10^{40} \, \rm erg/s$).
On the other hand, the hard X-ray luminosity of IHS-654 would be dominated by HMXBs.

Our analysis shows that systems hosting rapidly growing BH seeds at their center (such as  IHS-2538 and 
ILS-5836) can be easily discriminated from systems where the BH is almost inactive and/or the dominant 
emission comes from the host galaxy (such as ILS-654 and IHS-13356). 
They show bluer colors, intermediate IRX values and a high luminosity
in the hard X-ray band, largely dominated by the accreting BH. 
On the other hand, 
the discrimination of the nature of the BH seed, whether it formed as a 
Pop~III stellar remnant or as a DCBH, can be done only on the basis of statistical arguments. 

\section{Comparison with previous works}

In their recent works,  \citet{Pacucci15} and \citet{Natarajan17} 
suggest that JWST will be also able to discriminate among DCBH and
Pop~III remnant seeds by means of their spectral features.
Similarly, we have explored the SED of 
growing DCBHs and Pop~III remnant BHs, comparing the expected spectral 
features of the two different channels, in the same spirit of these earlier works
but with some important differences.

{
The post-processing with Cloudy that we perform in our study is complementary to 
the one adopted by \citet{Natarajan17}. They consider an idealized set-up of a halo at a given 
redshift (with a given metallicity) while in our model the galaxy properties (e.g. the metallicity 
and dust-to-gas ratio) are derived from the cosmological evolution of the halos.
}

The adopted 
initial gas density profile in the model of \citet{Pacucci15} results in SED with more prominent emission 
lines (mainly H and He recombination lines). 
In our simulation, similar gas densities and IR spectral features
are found in halos 13356 and 5836, in the very early stages of their evolution, at $\rm ages < 5-20$ Myr,
when - however -  AGN are faint $L_{\rm bol}\leq 10^{43}-10^{42}$, respectively. 

On the other hand, the systems investigated by \citet{Natarajan17} are the peculiar, transient OBGs 
\citep[][]{Agarwal13}.
These are different from our IHS and ILS hosts, from the evolutionary point of view.
In the scenario predicted by \citet{Agarwal13}, the newly formed DCBH host merge with a nearby star forming galaxy 
soon after the formation of the seed, in less than $\sim 1$ Myr. 
We find that, in a comparable time-scale, star formation in IHS hosts is enabled by the fast metal enrichment 
driven by gas-rich inflows from the IGM.
Systems similar to IHS-2538, the most massive accreting heavy seed, at the end of their isolated evolution
have properties consistent with an OBG, in terms of stellar and BH mass\footnote{For example, after living 
in isolation for 250 Myr, at $z=10$, the IHS-2538 host galaxy merges with a normal star-forming galaxy 
(not hosting a BH in its center) with a stellar mass of $\sim 3\times 10^7 \, M_\odot$ (before the merger). 
The properties of the galaxy formed by this merger are very similar to those expected for the 
OBGs discussed by \citet{Agarwal13}: the BH mass is $\sim 2\times 10^7 \, M_\odot$ and the 
stellar mass is $\sim 4\times 10^7 \, M_\odot$.}.

{
If compared with previous works, our model shows how the emission depends on the environmental 
conditions in which the BH seeds form and grow. These conditions are determined by the 
assembly history of the $z\sim  6$ SMBH and its host galaxy\footnote{The assembly history determines the 
interplay between different physical processes (e.g. Pop~III/Pop~II star formation, onset of DCBH seeding 
conditions, BH growth, build up of UV background etc.) and their effects on the environment (via chemical, 
mechanical, and radiative feedback).}. 
Our predictions are for sources that would be seen in biased (over-dense) regions where $z>6$ SMBH form
and, thus, are somehow complementary with respect to other works. 

In over-dense regions, the evolution of massive seeds (DCBHs) may be faster (more efficient) than in average 
volumes of the Universe (unbiased regions).
In the latter case, the ISM evolution is strongly dominated by star formation, where SN feedback regulates 
the growth (and the feedback strength) of BH seeds in low-mass galaxies  
at early cosmic epochs 
\citep[e.g.][]{Dubois15, Habouzit17, Bower17, AnglesAlcazar17, Trebitsch17}. 
In this scenario, growing BH seeds are similar to IHS-13356.

Conversely, in biased regions were $z>6$ SMBH are expected to form, the early growth of some massive BH seeds 
($\sim 33\%$ of the IHS population in our model) 
is self-regulated by AGN feedback, as star formation 
(and thus SN feedback) is quenched as a consequence of the rapid build up of an intense UV background radiation
(see Appendix A, paperI and V16 for details).
The comparison with other studies, thus, points to a different evolution in biased (over dense) and unbiased (average 
volumes) regions. This is consistent with the discussion presented in paperI.

In deep blank-field surveys, we expect that sources predicted in our model, would be seen in over-dense regions 
that will later form $z\sim 6$ quasars \citep[i.e. protoclusters, e.g.][]{Overzier16}.
}

{We find that a color-color selection, similar to that proposed by \citet{Natarajan17}, would help in 
distinguishing AGN-dominated galaxies from starbursts.}
Although they have similar AB magnitudes ($M_{\rm AB}\leq 26$ in the NIRcam bands), ILS-654-like starbursts
are expected to be redder than OBGs and IHS-2538-like systems.
On the other hand, despite the lower AB magnitude\footnote{ILS-5836 reaches $\rm M_{AB}\sim 29$ only at 
the end of its isolated evolution, at $z\sim 12$, and would be detectable with NIRcam at lower redshifts.} 
rapidly growing ILS-5836 like systems have colors similar to IHS-2538 like ones, comparable to those 
expected for $z\sim 11$ OBGs. 
Thus, we find that a color-color selection is capable of uniquely identifying the nature of BH seeds
only in a statistical sense. In fact, as we will show in the next section, rapidly growing BH seeds 
that originate from DCBHs have a higher probability of being detected, in agreement with previous
findings \citep{Pacucci15, Natarajan17}.

Another source of confusion in tracing the observational features of the first accreting BHs has been 
recently pointed out by \citet{Volonteri17}. The authors show that in high redshift galaxies, 
hosting $<1$ Gyr old stellar populations, may be very difficult to distinguish stellar and AGN emission at 
rest frame UV/optical wavelengths.
Moreover, the emission from low-mass AGN (powered by $10^5-10^6 \, M_\odot$ BHs), and from HMXB at 
$[2-10]$ keV would be difficult to disentangle at $z>6$. 
The observational strategy we have suggested above would help in overcoming such a confusion problem.

\section{Summary and Conclusions}
\label{sec:conclusions}

In paperI we have investigated the physical properties of galaxies hosting light and heavy 
seeds in a cosmological context, following the formation and evolution of the first stars and BHs 
along the hierarchical assembly of a $z=6.4$ quasar like SDSS J1148+5251. 
In this work we have used those properties to model the SEDs of both accreting BHs and their 
stellar counterparts, taking into account 
their redshift/time evolution as well as the effect of the interplay between, 
chemical, mechanical and radiative feedback on the host galaxies.

To this aim, we have selected four interesting objects from our sample of accreting BHs, 
two of which representative of the high-mass end of their population distributions,
and two in the low-mass end, to emphasize differences and similarities between the two classes.
We have extracted their 
evolutionary properties as long as their host galaxies evolve in isolation, only through mass exchanges 
with the external medium (infall/outflows), with no mergers.
Still, the evolution of these objects is influenced by (and influence) the evolution of the other 
galaxies polluting the external medium with metals and dust and permeating the virtual volume 
in which they reside with photo-dissociating and ionizing radiation.

For the selected objects we modeled the intrinsic emission coming from the accreting BHs
and the stars. We then used these templates to predict the SED emerging from the host galaxy by
processing the intrinsic emission using the software \cloudy.
The information required to compute both the intrinsic and reprocessed emission, 
namely the Pop~III remnant BH mass, BH accretion rate, and the properties of their host galaxies, 
such as the time evolving gas density, metallicity and dust-to-gas ratio 
were all consistently computed by the model \gamete.

Our study  attempts to answer two important questions:
\textit {(i) Can we detect the (growing) BH seed progenitors of the first SMBHs?} and if 
yes, (ii) \textit{Can we identify the nature of BH seeds?}

To answer the first question,
we confirm that future, high sensitivity facilities and missions, like JWST and ATHENA will be able to 
detect the most luminous ($L_{\rm NIR}\geq 10^{42} \, \rm erg/s$; $L_{\rm MIR}\geq 10^{43} \, \rm erg/s$; 
$L_{\rm [0.5-2] \, \rm keV}\geq 10^{42}$ erg s$^{-1}$) progenitors of the first SMBHs. 

Regarding the second question, searches for growing BH seeds 
must be extended out to $z>10$, when the probability
of observing them in isolation is expected to be higher (see paperI). 
{It is also interesting to note that at $z\lesssim 15 (17)$, 
for IHS (ILS), at least half of the BHs are still in isolated systems.}
At lower redshift, BH seeds and their host galaxies 
rapidly loose memory of their initial conditions (see paperI for details).

\begin{figure}
\centering
\includegraphics [width=8cm]{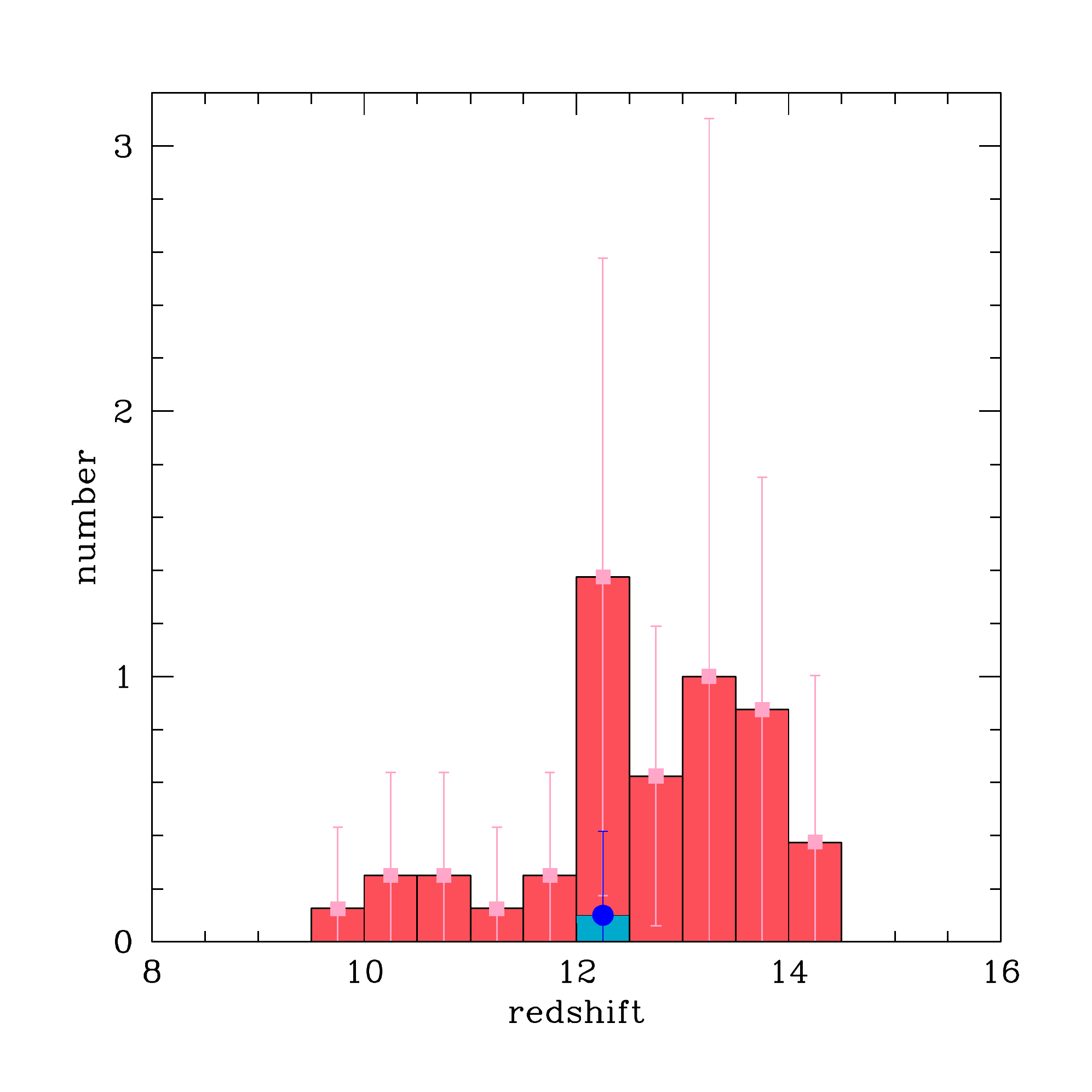}
\caption{The redshift distribution of detectable IHS (red) and ILS (blue). 
Histograms and data points show the average number over 10 different realizations 
with error bars indicating the $1\sigma$ dispersion. 
In each redshift bin, we show the number of systems with $L_{\rm bol}\geq 4\times 10^{43} \, \rm erg/s$ 
($M_{\rm BH}\geq 3.5\times 10^5 \, M_\odot$).
}
\label{fig:detectableSeeds} 
\end{figure}

{Among $z>6$ SMBH progenitors,
rapidly growing BH seeds at $z>10$ that are luminous enough to be detectable with JWST, in the field of
a quasar, are only $2\%$ of the total BH seed population. This percentage corresponds to an average number 
of $\sim 4^{+3}_{-2}$ (at a 1$\sigma$ confidence level, assuming Poissonian statistics) progenitors for 
each $z\sim 6$ SMBH. Their average redshift distribution is shown in 
Figure~\ref{fig:detectableSeeds}, where we also discriminate IHS (red histogram) and ILS (blue histogram).
Hence, a JWST detected BH progenitor at $z>10$ has a larger probability to originate from a DCBH than
from a Pop~III remnant.
Our conclusion is that it will be difficult to discriminate the nature of $z\sim 6$ SMBHs seeds on the 
basis of their observational features as Pop~III BHs and DCBHs look very similar when they are above JWST
or ATHENA sensitivity limits. However, if detected, these systems are more likely powered by a growing DCBH.
}

Interestingly, with a 10 ks NIRcam observation, sources like IHS-13356 can be detected with a SNR$=10$
in the F200W and F277W filters, already at $z\sim 15.7$. 
In addition, we foresee that blind search programs conducted with the spectrograph NIRspec would allow 
to reveal these kind of systems even at earlier epochs ($z>16$), thanks to the strong emission lines 
emerging in the spectra. 

We believe that a multi-wavelength approach will be fundamental in searching the first SMBH progenitors.
While JWST will be able to characterize their host galaxies, the high ATHENA flux sensitivity
($\sim 10^{-17} \, \rm erg/s/cm^2$ at $[0.5-2]$ keV, \citealt{Nandra13})
will allow to uncover the signatures of accreting BHs. 
A IR color-color plus IRX-$\beta$ selection of high redshift JWST sources could help in successfully 
identify possible SMBH progenitors candidates in future X-ray surveys, also providing a unique tool 
for preparing follow-up observations. 

Finally, we stress that the above conclusions apply to a scenario where SMBHs at $z\sim 6$ originate from the 
Eddington-limited growth of Pop~III BHs and DCBHs. If Pop~III stellar remnants are able to sustain short and 
intermittent phases of super-Eddington accretion \citep[][]{P16}, their occurrence and observational properties 
may be different.

\section*{Acknowledgments}
The research leading to these results has received funding from the European Research Council 
under the European Union's Seventh Framework Programme (FP/2007-2013) / ERC Grant Agreement n. 306476. 
LZ acknowledges financial support under ASI/INAF contract $1/037/12/0$.
We thank Angela Bongiorno, Fabrizio Fiore, Emanuele Giallongo, Michele Ginolfi, Andrea Grazian, 
Tilman Hartwig, Elisabeta Lusso and Laura Pentericci for fruitful discussions.

\bibliography{biblioReview}

\begin{thebibliography}{}
\makeatletter
\relax
\def\mn@urlcharsother{\let\do\@makeother \do\$\do\&\do\#\do\^\do\_\do\%\do\~}
\def\mn@doi{\begingroup\mn@urlcharsother \@ifnextchar [ {\mn@doi@}
  {\mn@doi@[]}}
\def\mn@doi@[#1]#2{\def\@tempa{#1}\ifx\@tempa\@empty \href
  {http://dx.doi.org/#2} {doi:#2}\else \href {http://dx.doi.org/#2} {#1}\fi
  \endgroup}
\def\mn@eprint#1#2{\mn@eprint@#1:#2::\@nil}
\def\mn@eprint@arXiv#1{\href {http://arxiv.org/abs/#1} {{\tt arXiv:#1}}}
\def\mn@eprint@dblp#1{\href {http://dblp.uni-trier.de/rec/bibtex/#1.xml}
  {dblp:#1}}
\def\mn@eprint@#1:#2:#3:#4\@nil{\def\@tempa {#1}\def\@tempb {#2}\def\@tempc
  {#3}\ifx \@tempc \@empty \let \@tempc \@tempb \let \@tempb \@tempa \fi \ifx
  \@tempb \@empty \def\@tempb {arXiv}\fi \@ifundefined
  {mn@eprint@\@tempb}{\@tempb:\@tempc}{\expandafter \expandafter \csname
  mn@eprint@\@tempb\endcsname \expandafter{\@tempc}}}

\bibitem[\protect\citeauthoryear{{Agarwal}, {Davis}, {Khochfar}, {Natarajan}
  \& {Dunlop}}{{Agarwal} et~al.}{2013}]{Agarwal13}
{Agarwal} B.,  {Davis} A.~J.,  {Khochfar} S.,  {Natarajan} P.,   {Dunlop}
  J.~S.,  2013, \mn@doi [\mnras] {10.1093/mnras/stt696}, \href
  {http://adsabs.harvard.edu/abs/2013MNRAS.432.3438A} {432, 3438}

\bibitem[\protect\citeauthoryear{{Agarwal}, {Dalla Vecchia}, {Johnson},
  {Khochfar}  \& {Paardekooper}}{{Agarwal} et~al.}{2014}]{Agarwal14}
{Agarwal} B.,  {Dalla Vecchia} C.,  {Johnson} J.~L.,  {Khochfar} S.,
  {Paardekooper} J.-P.,  2014, \mn@doi [\mnras] {10.1093/mnras/stu1112}, \href
  {http://adsabs.harvard.edu/abs/2014MNRAS.443..648A} {443, 648}

\bibitem[\protect\citeauthoryear{{Agarwal}, {Johnson}, {Zackrisson}, {Labbe},
  {van den Bosch}, {Natarajan}  \& {Khochfar}}{{Agarwal}
  et~al.}{2016}]{Agarwal16b}
{Agarwal} B.,  {Johnson} J.~L.,  {Zackrisson} E.,  {Labbe} I.,  {van den Bosch}
  F.~C.,  {Natarajan} P.,   {Khochfar} S.,  2016, \mn@doi [\mnras]
  {10.1093/mnras/stw1173}, \href
  {http://adsabs.harvard.edu/abs/2016MNRAS.460.4003A} {460, 4003}

\bibitem[\protect\citeauthoryear{{Agarwal}, {Johnson}, {Khochfar},
  {Pellegrini}, {Rydberg}, {Klessen}  \& {Oesch}}{{Agarwal}
  et~al.}{2017}]{Agarwal17a}
{Agarwal} B.,  {Johnson} J.~L.,  {Khochfar} S.,  {Pellegrini} E.,  {Rydberg}
  C.-E.,  {Klessen} R.~S.,   {Oesch} P.,  2017, preprint, \href
  {http://adsabs.harvard.edu/abs/2017arXiv170200407A} {} (\mn@eprint {arXiv}
  {1702.00407})

\bibitem[\protect\citeauthoryear{{Aird}, {Comastri}, {Brusa}, {Cappelluti},
  {Moretti}  \& {Vanzella}}{{Aird} et~al.}{2013}]{Aird13}
{Aird} J.,  {Comastri} A.,  {Brusa} M.,  {Cappelluti} N.,  {Moretti} A.,
  {Vanzella} e.~a.,  2013, preprint, \href
  {http://adsabs.harvard.edu/abs/2013arXiv1306.2325A} {} (\mn@eprint {arXiv}
  {1306.2325})

\bibitem[\protect\citeauthoryear{{Angl{\'e}s-Alc{\'a}zar},
  {Faucher-Gigu{\`e}re}, {Quataert}, {Hopkins}, {Feldmann}, {Torrey}, {Wetzel}
  \& {Kere{\v s}}}{{Angl{\'e}s-Alc{\'a}zar} et~al.}{2017}]{AnglesAlcazar17}
{Angl{\'e}s-Alc{\'a}zar} D.,  {Faucher-Gigu{\`e}re} C.-A.,  {Quataert} E.,
  {Hopkins} P.~F.,  {Feldmann} R.,  {Torrey} P.,  {Wetzel} A.,   {Kere{\v s}}
  D.,  2017, \mn@doi [\mnras] {10.1093/mnrasl/slx161}, \href
  {http://adsabs.harvard.edu/abs/2017MNRAS.472L.109A} {472, L109}

\bibitem[\protect\citeauthoryear{{Barisic} et~al.,}{{Barisic}
  et~al.}{2017}]{Barisic17}
{Barisic} I.,  et~al., 2017, \mn@doi [\apj] {10.3847/1538-4357/aa7eda}, \href
  {http://adsabs.harvard.edu/abs/2017ApJ...845...41B} {845, 41}

\bibitem[\protect\citeauthoryear{{Begelman}, {Volonteri}  \& {Rees}}{{Begelman}
  et~al.}{2006}]{Begelman06}
{Begelman} M.~C.,  {Volonteri} M.,   {Rees} M.~J.,  2006, \mn@doi [\mnras]
  {10.1111/j.1365-2966.2006.10467.x}, \href
  {http://adsabs.harvard.edu/abs/2006MNRAS.370..289B} {370, 289}

\bibitem[\protect\citeauthoryear{{Bouwens} et~al.,}{{Bouwens}
  et~al.}{2010}]{Bouwens10}
{Bouwens} R.~J.,  et~al., 2010, \mn@doi [\apjl] {10.1088/2041-8205/708/2/L69},
  \href {http://adsabs.harvard.edu/abs/2010ApJ...708L..69B} {708, L69}

\bibitem[\protect\citeauthoryear{{Bower}, {Schaye}, {Frenk}, {Theuns},
  {Schaller}, {Crain}  \& {McAlpine}}{{Bower} et~al.}{2017}]{Bower17}
{Bower} R.~G.,  {Schaye} J.,  {Frenk} C.~S.,  {Theuns} T.,  {Schaller} M.,
  {Crain} R.~A.,   {McAlpine} S.,  2017, \mn@doi [\mnras]
  {10.1093/mnras/stw2735}, \href
  {http://adsabs.harvard.edu/abs/2017MNRAS.465...32B} {465, 32}

\bibitem[\protect\citeauthoryear{Bowler, McLure, Dunlop, McLeod, Stanway,
  Eldridge  \& Jarvis}{Bowler et~al.}{2016}]{Bowler16}
Bowler R. A.~A.,  McLure R.~J.,  Dunlop J.~S.,  McLeod D.~J.,  Stanway E.~R.,
  Eldridge J.~J.,   Jarvis M.~J.,  2016, arXiv.org, p. arXiv:1609.00727

\bibitem[\protect\citeauthoryear{{Brightman}, {Silverman}, {Mainieri}, {Ueda},
  {Schramm}  \& et al.}{{Brightman} et~al.}{2013}]{Brightman2013}
{Brightman} M.,  {Silverman} J.~D.,  {Mainieri} V.,  {Ueda} Y.,  {Schramm} M.,
   et al. M.,  2013, \mn@doi [\mnras] {10.1093/mnras/stt920}, \href
  {http://adsabs.harvard.edu/abs/2013MNRAS.433.2485B} {433, 2485}

\bibitem[\protect\citeauthoryear{{Bromm} \& {Loeb}}{{Bromm} \&
  {Loeb}}{2003}]{BL03}
{Bromm} V.,  {Loeb} A.,  2003, \mn@doi [\apj] {10.1086/377529}, \href
  {http://adsabs.harvard.edu/abs/2003ApJ...596...34B} {596, 34}

\bibitem[\protect\citeauthoryear{{Burlon}, {Ajello}, {Greiner}, {Comastri},
  {Merloni}  \& {Gehrels}}{{Burlon} et~al.}{2011}]{Burlon11}
{Burlon} D.,  {Ajello} M.,  {Greiner} J.,  {Comastri} A.,  {Merloni} A.,
  {Gehrels} N.,  2011, \mn@doi [\apj] {10.1088/0004-637X/728/1/58}, \href
  {http://adsabs.harvard.edu/abs/2011ApJ...728...58B} {728, 58}

\bibitem[\protect\citeauthoryear{{Cappelluti}, {Comastri}, {Fontana},
  {Zamorani}, {Amorin}, {Castellano}  \& {Merlin}}{{Cappelluti}
  et~al.}{2016}]{Cappelluti16}
{Cappelluti} N.,  {Comastri} A.,  {Fontana} A.,  {Zamorani} G.,  {Amorin} R.,
  {Castellano} M.,   {Merlin} E. e.~a.,  2016, \mn@doi [\apj]
  {10.3847/0004-637X/823/2/95}, \href
  {http://adsabs.harvard.edu/abs/2016ApJ...823...95C} {823, 95}

\bibitem[\protect\citeauthoryear{{Chon}, {Hirano}, {Hosokawa}  \&
  {Yoshida}}{{Chon} et~al.}{2016}]{Chon16}
{Chon} S.,  {Hirano} S.,  {Hosokawa} T.,   {Yoshida} N.,  2016, \mn@doi [\apj]
  {10.3847/0004-637X/832/2/134}, \href
  {http://adsabs.harvard.edu/abs/2016ApJ...832..134C} {832, 134}

\bibitem[\protect\citeauthoryear{{Cicone} et~al.,}{{Cicone}
  et~al.}{2015}]{Cicone15}
{Cicone} C.,  et~al., 2015, \mn@doi [\aap] {10.1051/0004-6361/201424980}, \href
  {http://adsabs.harvard.edu/abs/2015A%26A...574A..14C} {574, A14}

\bibitem[\protect\citeauthoryear{{Cowie} \& {Songaila}}{{Cowie} \&
  {Songaila}}{1986}]{CowieSongaila86}
{Cowie} L.~L.,  {Songaila} A.,  1986, \mn@doi [\araa]
  {10.1146/annurev.aa.24.090186.002435}, \href
  {http://adsabs.harvard.edu/abs/1986ARA%26A..24..499C} {24, 499}

\bibitem[\protect\citeauthoryear{{Davies} et~al.,}{{Davies}
  et~al.}{2006}]{Davies06}
{Davies} R.~I.,  et~al., 2006, \mn@doi [\apj] {10.1086/504963}, \href
  {http://adsabs.harvard.edu/abs/2006ApJ...646..754D} {646, 754}

\bibitem[\protect\citeauthoryear{{Davies}, {M{\"u}ller S{\'a}nchez}, {Genzel},
  {Tacconi}, {Hicks}, {Friedrich}  \& {Sternberg}}{{Davies}
  et~al.}{2007}]{Davies07}
{Davies} R.~I.,  {M{\"u}ller S{\'a}nchez} F.,  {Genzel} R.,  {Tacconi} L.~J.,
  {Hicks} E.~K.~S.,  {Friedrich} S.,   {Sternberg} A.,  2007, \mn@doi [\apj]
  {10.1086/523032}, \href {http://adsabs.harvard.edu/abs/2007ApJ...671.1388D}
  {671, 1388}

\bibitem[\protect\citeauthoryear{{Di Matteo}, {Springel}  \& {Hernquist}}{{Di
  Matteo} et~al.}{2005}]{DiMatteo05}
{Di Matteo} T.,  {Springel} V.,   {Hernquist} L.,  2005, \mn@doi [\nat]
  {10.1038/nature03335}, \href
  {http://adsabs.harvard.edu/abs/2005Natur.433..604D} {433, 604}

\bibitem[\protect\citeauthoryear{{Dijkstra}, {Ferrara}  \&
  {Mesinger}}{{Dijkstra} et~al.}{2014}]{Dijkstra14}
{Dijkstra} M.,  {Ferrara} A.,   {Mesinger} A.,  2014, \mn@doi [\mnras]
  {10.1093/mnras/stu1007}, \href
  {http://adsabs.harvard.edu/abs/2014MNRAS.442.2036D} {442, 2036}

\bibitem[\protect\citeauthoryear{{Dijkstra}, {Gronke}  \& {Sobral}}{{Dijkstra}
  et~al.}{2016}]{Dijkstra16}
{Dijkstra} M.,  {Gronke} M.,   {Sobral} D.,  2016, \mn@doi [\apj]
  {10.3847/0004-637X/823/2/74}, \href
  {http://adsabs.harvard.edu/abs/2016ApJ...823...74D} {823, 74}

\bibitem[\protect\citeauthoryear{{Dubois}, {Volonteri}, {Silk}, {Devriendt},
  {Slyz}  \& {Teyssier}}{{Dubois} et~al.}{2015}]{Dubois15}
{Dubois} Y.,  {Volonteri} M.,  {Silk} J.,  {Devriendt} J.,  {Slyz} A.,
  {Teyssier} R.,  2015, \mn@doi [\mnras] {10.1093/mnras/stv1416}, \href
  {http://adsabs.harvard.edu/abs/2015MNRAS.452.1502D} {452, 1502}

\bibitem[\protect\citeauthoryear{{Ferland} et~al.,}{{Ferland}
  et~al.}{2013}]{Ferland13}
{Ferland} G.~J.,  et~al., 2013, \rmxaa, \href
  {http://adsabs.harvard.edu/abs/2013RMxAA..49..137F} {49, 137}

\bibitem[\protect\citeauthoryear{{Ferrara}, {Salvadori}, {Yue}  \&
  {Schleicher}}{{Ferrara} et~al.}{2014}]{Ferrara14}
{Ferrara} A.,  {Salvadori} S.,  {Yue} B.,   {Schleicher} D.,  2014, \mn@doi
  [\mnras] {10.1093/mnras/stu1280}, \href
  {http://adsabs.harvard.edu/abs/2014MNRAS.443.2410F} {443, 2410}

\bibitem[\protect\citeauthoryear{{Fioc} \& {Rocca-Volmerange}}{{Fioc} \&
  {Rocca-Volmerange}}{1997}]{Fioc97}
{Fioc} M.,  {Rocca-Volmerange} B.,  1997, \aap, \href
  {http://adsabs.harvard.edu/abs/1997A%26A...326..950F} {326, 950}

\bibitem[\protect\citeauthoryear{{Fioc} \& {Rocca-Volmerange}}{{Fioc} \&
  {Rocca-Volmerange}}{1999}]{Fioc99}
{Fioc} M.,  {Rocca-Volmerange} B.,  1999, ArXiv Astrophysics e-prints, \href
  {http://adsabs.harvard.edu/abs/1999astro.ph.12179F} {}

\bibitem[\protect\citeauthoryear{{Gallerani}, {Fan}, {Maiolino}  \&
  {Pacucci}}{{Gallerani} et~al.}{2017}]{Gallerani17}
{Gallerani} S.,  {Fan} X.,  {Maiolino} R.,   {Pacucci} F.,  2017, \mn@doi
  [\pasa] {10.1017/pasa.2017.14}, \href
  {http://adsabs.harvard.edu/abs/2017PASA...34...22G} {34, e022}

\bibitem[\protect\citeauthoryear{{Georgakakis} et~al.,}{{Georgakakis}
  et~al.}{2015}]{Georgakakis15}
{Georgakakis} A.,  et~al., 2015, \mn@doi [\mnras] {10.1093/mnras/stv1703},
  \href {http://adsabs.harvard.edu/abs/2015MNRAS.453.1946G} {453, 1946}

\bibitem[\protect\citeauthoryear{{Giallongo}, {Grazian}, {Fiore}, {Fontana},
  {Pentericci}  \& {Vanzella}}{{Giallongo} et~al.}{2015}]{Giallongo15}
{Giallongo} E.,  {Grazian} A.,  {Fiore} F.,  {Fontana} A.,  {Pentericci} L.,
  {Vanzella} E. e.~a.,  2015, \mn@doi [\aap] {10.1051/0004-6361/201425334},
  \href {http://adsabs.harvard.edu/abs/2015A\%26A...578A..83G} {578, A83}

\bibitem[\protect\citeauthoryear{{Habouzit} et~al.,}{{Habouzit}
  et~al.}{2016}]{Habouzit16hSAM}
{Habouzit} M.,  et~al., 2016, \mn@doi [\mnras] {10.1093/mnras/stv2740}, \href
  {http://adsabs.harvard.edu/abs/2016MNRAS.456.1901H} {456, 1901}

\bibitem[\protect\citeauthoryear{{Habouzit}, {Volonteri}  \&
  {Dubois}}{{Habouzit} et~al.}{2017}]{Habouzit17}
{Habouzit} M.,  {Volonteri} M.,   {Dubois} Y.,  2017, \mn@doi [\mnras]
  {10.1093/mnras/stx666}, \href
  {http://adsabs.harvard.edu/abs/2017MNRAS.468.3935H} {468, 3935}

\bibitem[\protect\citeauthoryear{{Haiman}}{{Haiman}}{2004}]{Haiman04}
{Haiman} Z.,  2004, \mn@doi [\apj] {10.1086/422910}, \href
  {http://adsabs.harvard.edu/abs/2004ApJ...613...36H} {613, 36}

\bibitem[\protect\citeauthoryear{{Hartwig}, {Glover}, {Klessen}, {Latif}  \&
  {Volonteri}}{{Hartwig} et~al.}{2015}]{Hartwig15}
{Hartwig} T.,  {Glover} S.~C.~O.,  {Klessen} R.~S.,  {Latif} M.~A.,
  {Volonteri} M.,  2015, \mn@doi [\mnras] {10.1093/mnras/stv1368}, \href
  {http://adsabs.harvard.edu/abs/2015MNRAS.452.1233H} {452, 1233}

\bibitem[\protect\citeauthoryear{{Heger}, {Fryer}, {Woosley}, {Langer}  \&
  {Hartmann}}{{Heger} et~al.}{2003}]{Heger03}
{Heger} A.,  {Fryer} C.~L.,  {Woosley} S.~E.,  {Langer} N.,   {Hartmann} D.~H.,
   2003, \mn@doi [\apj] {10.1086/375341}, \href
  {http://adsabs.harvard.edu/abs/2003ApJ...591..288H} {591, 288}

\bibitem[\protect\citeauthoryear{{Hicks}, {Davies}, {Malkan}, {Genzel},
  {Tacconi}, {M{\"u}ller S{\'a}nchez}  \& {Sternberg}}{{Hicks}
  et~al.}{2009}]{Hicks09}
{Hicks} E.~K.~S.,  {Davies} R.~I.,  {Malkan} M.~A.,  {Genzel} R.,  {Tacconi}
  L.~J.,  {M{\"u}ller S{\'a}nchez} F.,   {Sternberg} A.,  2009, \mn@doi [\apj]
  {10.1088/0004-637X/696/1/448}, \href
  {http://adsabs.harvard.edu/abs/2009ApJ...696..448H} {696, 448}

\bibitem[\protect\citeauthoryear{{Hirano}, {Hosokawa}, {Yoshida}, {Omukai}  \&
  {Yorke}}{{Hirano} et~al.}{2015}]{Hirano15}
{Hirano} S.,  {Hosokawa} T.,  {Yoshida} N.,  {Omukai} K.,   {Yorke} H.~W.,
  2015, \mn@doi [\mnras] {10.1093/mnras/stv044}, \href
  {http://adsabs.harvard.edu/abs/2015MNRAS.448..568H} {448, 568}

\bibitem[\protect\citeauthoryear{{Hosokawa}, {Omukai}  \& {Yorke}}{{Hosokawa}
  et~al.}{2012}]{Hosokawa12}
{Hosokawa} T.,  {Omukai} K.,   {Yorke} H.~W.,  2012, \mn@doi [\apj]
  {10.1088/0004-637X/756/1/93}, \href
  {http://adsabs.harvard.edu/abs/2012ApJ...756...93H} {756, 93}

\bibitem[\protect\citeauthoryear{{Hosokawa}, {Yorke}, {Inayoshi}, {Omukai}  \&
  {Yoshida}}{{Hosokawa} et~al.}{2013}]{Hosokawa13}
{Hosokawa} T.,  {Yorke} H.~W.,  {Inayoshi} K.,  {Omukai} K.,   {Yoshida} N.,
  2013, \mn@doi [\apj] {10.1088/0004-637X/778/2/178}, \href
  {http://adsabs.harvard.edu/abs/2013ApJ...778..178H} {778, 178}

\bibitem[\protect\citeauthoryear{{Inayoshi} \& {Haiman}}{{Inayoshi} \&
  {Haiman}}{2014}]{IH14}
{Inayoshi} K.,  {Haiman} Z.,  2014, \mn@doi [\mnras] {10.1093/mnras/stu1870},
  \href {http://adsabs.harvard.edu/abs/2014MNRAS.445.1549I} {445, 1549}

\bibitem[\protect\citeauthoryear{{Inayoshi} \& {Omukai}}{{Inayoshi} \&
  {Omukai}}{2012}]{IO12}
{Inayoshi} K.,  {Omukai} K.,  2012, \mn@doi [\mnras]
  {10.1111/j.1365-2966.2012.20812.x}, \href
  {http://adsabs.harvard.edu/abs/2012MNRAS.422.2539I} {422, 2539}

\bibitem[\protect\citeauthoryear{{Inayoshi}, {Omukai}  \& {Tasker}}{{Inayoshi}
  et~al.}{2014}]{Inayoshi14}
{Inayoshi} K.,  {Omukai} K.,   {Tasker} E.,  2014, \mn@doi [\mnras]
  {10.1093/mnrasl/slu151}, \href
  {http://adsabs.harvard.edu/abs/2014MNRAS.445L.109I} {445, L109}

\bibitem[\protect\citeauthoryear{{Inayoshi}, {Visbal}  \&
  {Kashiyama}}{{Inayoshi} et~al.}{2015}]{Inayoshi15}
{Inayoshi} K.,  {Visbal} E.,   {Kashiyama} K.,  2015, \mn@doi [\mnras]
  {10.1093/mnras/stv1654}, \href
  {http://adsabs.harvard.edu/abs/2015MNRAS.453.1692I} {453, 1692}

\bibitem[\protect\citeauthoryear{{Inoue}, {Shimizu}, {Iwata}  \&
  {Tanaka}}{{Inoue} et~al.}{2014}]{Inoue14}
{Inoue} A.~K.,  {Shimizu} I.,  {Iwata} I.,   {Tanaka} M.,  2014, \mn@doi
  [\mnras] {10.1093/mnras/stu936}, \href
  {http://adsabs.harvard.edu/abs/2014MNRAS.442.1805I} {442, 1805}

\bibitem[\protect\citeauthoryear{{Johnson} \& {Haardt}}{{Johnson} \&
  {Haardt}}{2016}]{JohnsonHaardt16}
{Johnson} J.~L.,  {Haardt} F.,  2016, \mn@doi [\pasa] {10.1017/pasa.2016.4},
  \href {http://adsabs.harvard.edu/abs/2016PASA...33....7J} {33, e007}

\bibitem[\protect\citeauthoryear{{Larson}}{{Larson}}{1998}]{Larson98}
{Larson} R.~B.,  1998, \mnras, 301, 569

\bibitem[\protect\citeauthoryear{{Latif} \& {Ferrara}}{{Latif} \&
  {Ferrara}}{2016}]{LatifFerrara16}
{Latif} M.~A.,  {Ferrara} A.,  2016, \mn@doi [\pasa] {10.1017/pasa.2016.41},
  \href {http://adsabs.harvard.edu/abs/2016PASA...33...51L} {33, e051}

\bibitem[\protect\citeauthoryear{{Latif}, {Schleicher}, {Schmidt}  \&
  {Niemeyer}}{{Latif} et~al.}{2013}]{Latif13b}
{Latif} M.~A.,  {Schleicher} D.~R.~G.,  {Schmidt} W.,   {Niemeyer} J.~C.,
  2013, \mn@doi [\mnras] {10.1093/mnras/stt1786}, \href
  {http://adsabs.harvard.edu/abs/2013MNRAS.436.2989L} {436, 2989}

\bibitem[\protect\citeauthoryear{{Lehmer}, {Basu-Zych}, {Mineo}, {Brandt},
  {Eufrasio}  \& {Fragos}}{{Lehmer} et~al.}{2016}]{Lehmer16}
{Lehmer} B.~D.,  {Basu-Zych} A.~R.,  {Mineo} S.,  {Brandt} W.~N.,  {Eufrasio}
  R.~T.,   {Fragos} e.~a.,  2016, \mn@doi [\apj] {10.3847/0004-637X/825/1/7},
  \href {http://adsabs.harvard.edu/abs/2016ApJ...825....7L} {825, 7}

\bibitem[\protect\citeauthoryear{{Lodato} \& {Natarajan}}{{Lodato} \&
  {Natarajan}}{2006}]{LodatoNatarajan06}
{Lodato} G.,  {Natarajan} P.,  2006, \mn@doi [\mnras]
  {10.1111/j.1365-2966.2006.10801.x}, \href
  {http://adsabs.harvard.edu/abs/2006MNRAS.371.1813L} {371, 1813}

\bibitem[\protect\citeauthoryear{{Lodato} \& {Natarajan}}{{Lodato} \&
  {Natarajan}}{2007}]{LodatoNatarajan07}
{Lodato} G.,  {Natarajan} P.,  2007, \mn@doi [\mnras]
  {10.1111/j.1745-3933.2007.00304.x}, \href
  {http://adsabs.harvard.edu/abs/2007MNRAS.377L..64L} {377, L64}

\bibitem[\protect\citeauthoryear{{Luo} et~al.,}{{Luo} et~al.}{2017}]{Luo17}
{Luo} B.,  et~al., 2017, \mn@doi [\apjs] {10.3847/1538-4365/228/1/2}, \href
  {http://adsabs.harvard.edu/abs/2017ApJS..228....2L} {228, 2}

\bibitem[\protect\citeauthoryear{{Madau}}{{Madau}}{1995}]{Madau95}
{Madau} P.,  1995, \mn@doi [\apj] {10.1086/175332}, \href
  {http://adsabs.harvard.edu/abs/1995ApJ...441...18M} {441, 18}

\bibitem[\protect\citeauthoryear{{Madau} \& {Rees}}{{Madau} \&
  {Rees}}{2001}]{MadauRees01}
{Madau} P.,  {Rees} M.~J.,  2001, \mn@doi [\apjl] {10.1086/319848}, \href
  {http://adsabs.harvard.edu/abs/2001ApJ...551L..27M} {551, L27}

\bibitem[\protect\citeauthoryear{{Madau}, {Rees}, {Volonteri}, {Haardt}  \&
  {Oh}}{{Madau} et~al.}{2004}]{Madau04}
{Madau} P.,  {Rees} M.~J.,  {Volonteri} M.,  {Haardt} F.,   {Oh} S.~P.,  2004,
  \mn@doi [\apj] {10.1086/381935}, \href
  {http://adsabs.harvard.edu/abs/2004ApJ...604..484M} {604, 484}

\bibitem[\protect\citeauthoryear{{Magdziarz} \& {Zdziarski}}{{Magdziarz} \&
  {Zdziarski}}{1995}]{Magdziarz95}
{Magdziarz} P.,  {Zdziarski} A.~A.,  1995, \mn@doi [\mnras]
  {10.1093/mnras/273.3.837}, \href
  {http://adsabs.harvard.edu/abs/1995MNRAS.273..837M} {273, 837}

\bibitem[\protect\citeauthoryear{{Maiolino} et~al.,}{{Maiolino}
  et~al.}{2012}]{Maiolino12}
{Maiolino} R.,  et~al., 2012, \mn@doi [\mnras]
  {10.1111/j.1745-3933.2012.01303.x}, \href
  {http://adsabs.harvard.edu/abs/2012MNRAS.425L..66M} {425, L66}

\bibitem[\protect\citeauthoryear{{Mancini}, {Schneider}, {Graziani},
  {Valiante}, {Dayal}, {Maio}  \& {Ciardi}}{{Mancini} et~al.}{2016}]{Mancini16}
{Mancini} M.,  {Schneider} R.,  {Graziani} L.,  {Valiante} R.,  {Dayal} P.,
  {Maio} U.,   {Ciardi} B.,  2016, \mn@doi [\mnras] {10.1093/mnras/stw1783},
  \href {http://adsabs.harvard.edu/abs/2016MNRAS.462.3130M} {462, 3130}

\bibitem[\protect\citeauthoryear{{Mather}, {Fixsen}, {Shafer}, {Mosier}  \&
  {Wilkinson}}{{Mather} et~al.}{1999}]{Mather99}
{Mather} J.~C.,  {Fixsen} D.~J.,  {Shafer} R.~A.,  {Mosier} C.,   {Wilkinson}
  D.~T.,  1999, \mn@doi [\apj] {10.1086/306805}, \href
  {http://adsabs.harvard.edu/abs/1999ApJ...512..511M} {512, 511}

\bibitem[\protect\citeauthoryear{{Mathis}, {Rumpl}  \& {Nordsieck}}{{Mathis}
  et~al.}{1977}]{Mathis77}
{Mathis} J.~S.,  {Rumpl} W.,   {Nordsieck} K.~H.,  1977, \mn@doi [\apj]
  {10.1086/155591}, \href {http://adsabs.harvard.edu/abs/1977ApJ...217..425M}
  {217, 425}

\bibitem[\protect\citeauthoryear{{Matthee}, {Sobral}, {Santos},
  {R{\"o}ttgering}, {Darvish}  \& {Mobasher}}{{Matthee}
  et~al.}{2015}]{Matthee15}
{Matthee} J.,  {Sobral} D.,  {Santos} S.,  {R{\"o}ttgering} H.,  {Darvish} B.,
   {Mobasher} B.,  2015, \mn@doi [\mnras] {10.1093/mnras/stv947}, \href
  {http://adsabs.harvard.edu/abs/2015MNRAS.451..400M} {451, 400}

\bibitem[\protect\citeauthoryear{{Mullman}, {Lawler}, {Zsarg{\'o}}  \&
  {Federman}}{{Mullman} et~al.}{1998}]{Mullman98}
{Mullman} K.~L.,  {Lawler} J.~E.,  {Zsarg{\'o}} J.,   {Federman} S.~R.,  1998,
  \mn@doi [\apj] {10.1086/305780}, \href
  {http://adsabs.harvard.edu/abs/1998ApJ...500.1064M} {500, 1064}

\bibitem[\protect\citeauthoryear{{Nandra} et~al.,}{{Nandra}
  et~al.}{2013}]{Nandra13}
{Nandra} K.,  et~al., 2013, preprint, \href
  {http://adsabs.harvard.edu/abs/2013arXiv1306.2307N} {} (\mn@eprint {arXiv}
  {1306.2307})

\bibitem[\protect\citeauthoryear{{Natarajan}}{{Natarajan}}{2011}]{Natarajan11}
{Natarajan} P.,  2011, preprint, \href
  {http://adsabs.harvard.edu/abs/2011arXiv1105.4902N} {} (\mn@eprint {arXiv}
  {1105.4902})

\bibitem[\protect\citeauthoryear{{Natarajan}, {Pacucci}, {Ferrara}, {Agarwal},
  {Ricarte}, {Zackrisson}  \& {Cappelluti}}{{Natarajan}
  et~al.}{2017}]{Natarajan17}
{Natarajan} P.,  {Pacucci} F.,  {Ferrara} A.,  {Agarwal} B.,  {Ricarte} A.,
  {Zackrisson} E.,   {Cappelluti} N.,  2017, \mn@doi [\apj]
  {10.3847/1538-4357/aa6330}, \href
  {http://adsabs.harvard.edu/abs/2017ApJ...838..117N} {838, 117}

\bibitem[\protect\citeauthoryear{{Overzier}}{{Overzier}}{2016}]{Overzier16}
{Overzier} R.~A.,  2016, \mn@doi [\aapr] {10.1007/s00159-016-0100-3}, \href
  {http://adsabs.harvard.edu/abs/2016A%26ARv..24...14O} {24, 14}

\bibitem[\protect\citeauthoryear{{Pacucci}, {Ferrara}, {Volonteri}  \&
  {Dubus}}{{Pacucci} et~al.}{2015}]{Pacucci15}
{Pacucci} F.,  {Ferrara} A.,  {Volonteri} M.,   {Dubus} G.,  2015, \mn@doi
  [\mnras] {10.1093/mnras/stv2196}, \href
  {http://adsabs.harvard.edu/abs/2015MNRAS.454.3771P} {454, 3771}

\bibitem[\protect\citeauthoryear{{Pacucci}, {Ferrara}, {Grazian}, {Fiore},
  {Giallongo}  \& {Puccetti}}{{Pacucci} et~al.}{2016}]{Pacucci16}
{Pacucci} F.,  {Ferrara} A.,  {Grazian} A.,  {Fiore} F.,  {Giallongo} E.,
  {Puccetti} S.,  2016, \mn@doi [\mnras] {10.1093/mnras/stw725}, \href
  {http://adsabs.harvard.edu/abs/2016MNRAS.459.1432P} {459, 1432}

\bibitem[\protect\citeauthoryear{{Pacucci}, {Pallottini}, {Ferrara}  \&
  {Gallerani}}{{Pacucci} et~al.}{2017}]{Pacucci17}
{Pacucci} F.,  {Pallottini} A.,  {Ferrara} A.,   {Gallerani} S.,  2017, \mn@doi
  [\mnras] {10.1093/mnrasl/slx029}, \href
  {http://adsabs.harvard.edu/abs/2017MNRAS.468L..77P} {468, L77}

\bibitem[\protect\citeauthoryear{{Pallottini} et~al.,}{{Pallottini}
  et~al.}{2015}]{Pallottini15}
{Pallottini} A.,  et~al., 2015, \mn@doi [\mnras] {10.1093/mnras/stv1795}, \href
  {http://adsabs.harvard.edu/abs/2015MNRAS.453.2465P} {453, 2465}

\bibitem[\protect\citeauthoryear{{Peebles}}{{Peebles}}{1971}]{Peebles71}
{Peebles} P.~J.~E.,  1971, {Physical cosmology}

\bibitem[\protect\citeauthoryear{{Petri}, {Ferrara}  \& {Salvaterra}}{{Petri}
  et~al.}{2012}]{Petri12}
{Petri} A.,  {Ferrara} A.,   {Salvaterra} R.,  2012, \mn@doi [\mnras]
  {10.1111/j.1365-2966.2012.20743.x}, \href
  {http://adsabs.harvard.edu/abs/2012MNRAS.422.1690P} {422, 1690}

\bibitem[\protect\citeauthoryear{{Pezzulli}, {Valiante}  \&
  {Schneider}}{{Pezzulli} et~al.}{2016}]{P16}
{Pezzulli} E.,  {Valiante} R.,   {Schneider} R.,  2016, \mn@doi [\mnras]
  {10.1093/mnras/stw505}, \href
  {http://adsabs.harvard.edu/abs/2016MNRAS.458.3047P} {458, 3047}

\bibitem[\protect\citeauthoryear{{Pezzulli}, {Valiante}, {Orofino},
  {Schneider}, {Gallerani}  \& {Sbarrato}}{{Pezzulli} et~al.}{2017}]{P17a}
{Pezzulli} E.,  {Valiante} R.,  {Orofino} M.~C.,  {Schneider} R.,  {Gallerani}
  S.,   {Sbarrato} T.,  2017, \mn@doi [\mnras] {10.1093/mnras/stw3243}, \href
  {http://adsabs.harvard.edu/abs/2017MNRAS.466.2131P} {466, 2131}

\bibitem[\protect\citeauthoryear{{Reines} \& {Comastri}}{{Reines} \&
  {Comastri}}{2016}]{ReinesComastri16}
{Reines} A.~E.,  {Comastri} A.,  2016, \mn@doi [\pasa] {10.1017/pasa.2016.46},
  \href {http://adsabs.harvard.edu/abs/2016PASA...33...54R} {33, e054}

\bibitem[\protect\citeauthoryear{{Salvadori}, {Schneider}  \&
  {Ferrara}}{{Salvadori} et~al.}{2007}]{SS07}
{Salvadori} S.,  {Schneider} R.,   {Ferrara} A.,  2007, \mn@doi [\mnras]
  {10.1111/j.1365-2966.2007.12133.x}, \href
  {http://adsabs.harvard.edu/abs/2007MNRAS.381..647S} {381, 647}

\bibitem[\protect\citeauthoryear{{Salvadori}, {Ferrara}  \&
  {Schneider}}{{Salvadori} et~al.}{2008}]{SS08}
{Salvadori} S.,  {Ferrara} A.,   {Schneider} R.,  2008, \mn@doi [\mnras]
  {10.1111/j.1365-2966.2008.13035.x}, \href
  {http://adsabs.harvard.edu/abs/2008MNRAS.386..348S} {386, 348}

\bibitem[\protect\citeauthoryear{{Sani}, {Marconi}, {Hunt}  \&
  {Risaliti}}{{Sani} et~al.}{2011}]{Sani11}
{Sani} E.,  {Marconi} A.,  {Hunt} L.~K.,   {Risaliti} G.,  2011, \mn@doi
  [\mnras] {10.1111/j.1365-2966.2011.18229.x}, \href
  {http://adsabs.harvard.edu/abs/2011MNRAS.413.1479S} {413, 1479}

\bibitem[\protect\citeauthoryear{{Sani} et~al.,}{{Sani} et~al.}{2012}]{Sani12}
{Sani} E.,  et~al., 2012, \mn@doi [\mnras] {10.1111/j.1365-2966.2012.21333.x},
  \href {http://adsabs.harvard.edu/abs/2012MNRAS.424.1963S} {424, 1963}

\bibitem[\protect\citeauthoryear{{Savage} \& {Sembach}}{{Savage} \&
  {Sembach}}{1996}]{SavageSembach96}
{Savage} B.~D.,  {Sembach} K.~R.,  1996, \mn@doi [\araa]
  {10.1146/annurev.astro.34.1.279}, \href
  {http://adsabs.harvard.edu/abs/1996ARA%26A..34..279S} {34, 279}

\bibitem[\protect\citeauthoryear{{Sazonov}, {Ostriker}  \& {Sunyaev}}{{Sazonov}
  et~al.}{2004}]{SOS04}
{Sazonov} S.~Y.,  {Ostriker} J.~P.,   {Sunyaev} R.~A.,  2004, \mn@doi [\mnras]
  {10.1111/j.1365-2966.2004.07184.x}, \href
  {http://adsabs.harvard.edu/abs/2004MNRAS.347..144S} {347, 144}

\bibitem[\protect\citeauthoryear{{Schaerer}}{{Schaerer}}{2002}]{Schaerer02}
{Schaerer} D.,  2002, \mn@doi [\aap] {10.1051/0004-6361:20011619}, \href
  {http://adsabs.harvard.edu/abs/2002A%26A...382...28S} {382, 28}

\bibitem[\protect\citeauthoryear{{Schneider}, {Ferrara}, {Natarajan}  \&
  {Omukai}}{{Schneider} et~al.}{2002}]{Schneider02}
{Schneider} R.,  {Ferrara} A.,  {Natarajan} P.,   {Omukai} K.,  2002, \mn@doi
  [\apj] {10.1086/339917}, \href
  {http://adsabs.harvard.edu/abs/2002ApJ...571...30S} {571, 30}

\bibitem[\protect\citeauthoryear{{Schneider}, {Ferrara}, {Salvaterra}, {Omukai}
   \& {Bromm}}{{Schneider} et~al.}{2003}]{Schneider03}
{Schneider} R.,  {Ferrara} A.,  {Salvaterra} R.,  {Omukai} K.,   {Bromm} V.,
  2003, \nat, \href {http://adsabs.harvard.edu/abs/2003Natur.422..869S} {422,
  869}

\bibitem[\protect\citeauthoryear{{Schneider}, {Omukai}, {Bianchi}  \&
  {Valiante}}{{Schneider} et~al.}{2012}]{Schneider12}
{Schneider} R.,  {Omukai} K.,  {Bianchi} S.,   {Valiante} R.,  2012, \mn@doi
  [\mnras] {10.1111/j.1365-2966.2011.19818.x}, \href
  {http://adsabs.harvard.edu/abs/2012MNRAS.419.1566S} {419, 1566}

\bibitem[\protect\citeauthoryear{{Shakura} \& {Sunyaev}}{{Shakura} \&
  {Sunyaev}}{1973}]{SS1973}
{Shakura} N.~I.,  {Sunyaev} R.~A.,  1973, \aap, 24, 337

\bibitem[\protect\citeauthoryear{{Smidt}, {Wiggins}  \& {Johnson}}{{Smidt}
  et~al.}{2016}]{Smidt16}
{Smidt} J.,  {Wiggins} B.~K.,   {Johnson} J.~L.,  2016, \mn@doi [\apjl]
  {10.3847/2041-8205/829/1/L6}, \href
  {http://adsabs.harvard.edu/abs/2016ApJ...829L...6S} {829, L6}

\bibitem[\protect\citeauthoryear{{Smith}, {Bromm}  \& {Loeb}}{{Smith}
  et~al.}{2016}]{Smith16}
{Smith} A.,  {Bromm} V.,   {Loeb} A.,  2016, \mn@doi [\mnras]
  {10.1093/mnras/stw1129}, \href
  {http://adsabs.harvard.edu/abs/2016MNRAS.460.3143S} {460, 3143}

\bibitem[\protect\citeauthoryear{{Snow}, {Destree}  \& {Jensen}}{{Snow}
  et~al.}{2007}]{Snow07}
{Snow} T.~P.,  {Destree} J.~D.,   {Jensen} A.~G.,  2007, \mn@doi [\apj]
  {10.1086/510187}, \href {http://adsabs.harvard.edu/abs/2007ApJ...655..285S}
  {655, 285}

\bibitem[\protect\citeauthoryear{{Sobral}, {Matthee}, {Darvish}, {Schaerer},
  {Mobasher}, {R{\"o}ttgering}, {Santos}  \& {Hemmati}}{{Sobral}
  et~al.}{2015}]{Sobral15}
{Sobral} D.,  {Matthee} J.,  {Darvish} B.,  {Schaerer} D.,  {Mobasher} B.,
  {R{\"o}ttgering} H.~J.~A.,  {Santos} S.,   {Hemmati} S.,  2015, \mn@doi
  [\apj] {10.1088/0004-637X/808/2/139}, \href
  {http://adsabs.harvard.edu/abs/2015ApJ...808..139S} {808, 139}

\bibitem[\protect\citeauthoryear{{Sugimura}, {Coppola}, {Omukai}, {Galli}  \&
  {Palla}}{{Sugimura} et~al.}{2015}]{Sugimura15}
{Sugimura} K.,  {Coppola} C.~M.,  {Omukai} K.,  {Galli} D.,   {Palla} F.,
  2015, arXiv:1509.04562, \href
  {http://adsabs.harvard.edu/abs/2015arXiv150904562S} {}

\bibitem[\protect\citeauthoryear{{Tanaka} \& {Haiman}}{{Tanaka} \&
  {Haiman}}{2009}]{TH09}
{Tanaka} T.,  {Haiman} Z.,  2009, \mn@doi [\apj]
  {10.1088/0004-637X/696/2/1798}, \href
  {http://adsabs.harvard.edu/abs/2009ApJ...696.1798T} {696, 1798}

\bibitem[\protect\citeauthoryear{{Trebitsch}, {Volonteri}, {Dubois}  \&
  {Madau}}{{Trebitsch} et~al.}{2017}]{Trebitsch17}
{Trebitsch} M.,  {Volonteri} M.,  {Dubois} Y.,   {Madau} P.,  2017, preprint,
  \href {http://adsabs.harvard.edu/abs/2017arXiv171205804T} {} (\mn@eprint
  {arXiv} {1712.05804})

\bibitem[\protect\citeauthoryear{{Treister}, {Schawinski}, {Volonteri}  \&
  {Natarajan}}{{Treister} et~al.}{2013}]{Treister13}
{Treister} E.,  {Schawinski} K.,  {Volonteri} M.,   {Natarajan} P.,  2013,
  \mn@doi [\apj] {10.1088/0004-637X/778/2/130}, \href
  {http://adsabs.harvard.edu/abs/2013ApJ...778..130T} {778, 130}

\bibitem[\protect\citeauthoryear{{Valiante}, {Schneider}, {Salvadori}  \&
  {Bianchi}}{{Valiante} et~al.}{2011}]{V11}
{Valiante} R.,  {Schneider} R.,  {Salvadori} S.,   {Bianchi} S.,  2011, \mn@doi
  [\mnras] {10.1111/j.1365-2966.2011.19168.x}, \href
  {http://adsabs.harvard.edu/abs/2011MNRAS.416.1916V} {416, 1916}

\bibitem[\protect\citeauthoryear{{Valiante}, {Schneider}, {Maiolino},
  {Salvadori}  \& {Bianchi}}{{Valiante} et~al.}{2012}]{V12}
{Valiante} R.,  {Schneider} R.,  {Maiolino} R.,  {Salvadori} S.,   {Bianchi}
  S.,  2012, \mn@doi [\mnras] {10.1111/j.1745-3933.2012.01345.x}, \href
  {http://adsabs.harvard.edu/abs/2012MNRAS.427L..60V} {427, L60}

\bibitem[\protect\citeauthoryear{{Valiante}, {Schneider}, {Salvadori}  \&
  {Gallerani}}{{Valiante} et~al.}{2014}]{V14}
{Valiante} R.,  {Schneider} R.,  {Salvadori} S.,   {Gallerani} S.,  2014,
  \mn@doi [\mnras] {10.1093/mnras/stu1613}, \href
  {http://adsabs.harvard.edu/abs/2014MNRAS.444.2442V} {444, 2442}

\bibitem[\protect\citeauthoryear{{Valiante}, {Schneider}, {Volonteri}  \&
  {Omukai}}{{Valiante} et~al.}{2016}]{V16}
{Valiante} R.,  {Schneider} R.,  {Volonteri} M.,   {Omukai} K.,  2016, \mnras

\bibitem[\protect\citeauthoryear{{Valiante}, {Schneider}, {Graziani}  \&
  {Zappacosta}}{{Valiante} et~al.}{2017a}]{V17b}
{Valiante} R.,  {Schneider} R.,  {Graziani} L.,   {Zappacosta} L.,  2017a,
  preprint, \href {http://adsabs.harvard.edu/abs/2017arXiv171111033V} {}
  (\mn@eprint {arXiv} {1711.11033})

\bibitem[\protect\citeauthoryear{{Valiante}, {Agarwal}, {Habouzit}  \&
  {Pezzulli}}{{Valiante} et~al.}{2017b}]{V17a}
{Valiante} R.,  {Agarwal} B.,  {Habouzit} M.,   {Pezzulli} E.,  2017b, \mn@doi
  [\pasa] {10.1017/pasa.2017.25}, \href
  {http://adsabs.harvard.edu/abs/2017PASA...34...31V} {34, e031}

\bibitem[\protect\citeauthoryear{{Visbal}, {Haiman}  \& {Bryan}}{{Visbal}
  et~al.}{2016}]{Visbal16}
{Visbal} E.,  {Haiman} Z.,   {Bryan} G.~L.,  2016, \mn@doi [\mnras]
  {10.1093/mnrasl/slw071}, \href
  {http://adsabs.harvard.edu/abs/2016MNRAS.460L..59V} {460, L59}

\bibitem[\protect\citeauthoryear{{Vito} et~al.,}{{Vito} et~al.}{2016}]{Vito16}
{Vito} F.,  et~al., 2016, \mn@doi [\mnras] {10.1093/mnras/stw1998}, \href
  {http://adsabs.harvard.edu/abs/2016MNRAS.463..348V} {463, 348}

\bibitem[\protect\citeauthoryear{{Volonteri}}{{Volonteri}}{2010}]{Volonteri10}
{Volonteri} M.,  2010, \mn@doi [\araa] {10.1007/s00159-010-0029-x}, \href
  {http://adsabs.harvard.edu/abs/2010A%26ARv..18..279V} {18, 279}

\bibitem[\protect\citeauthoryear{{Volonteri} \& {Rees}}{{Volonteri} \&
  {Rees}}{2005}]{Volonteri05}
{Volonteri} M.,  {Rees} M.~J.,  2005, \mn@doi [\apj] {10.1086/466521}, \href
  {http://adsabs.harvard.edu/abs/2005ApJ...633..624V} {633, 624}

\bibitem[\protect\citeauthoryear{{Volonteri} \& {Rees}}{{Volonteri} \&
  {Rees}}{2006}]{Volonteri06}
{Volonteri} M.,  {Rees} M.~J.,  2006, \mn@doi [\apj] {10.1086/507444}, \href
  {http://adsabs.harvard.edu/abs/2006ApJ...650..669V} {650, 669}

\bibitem[\protect\citeauthoryear{{Volonteri}, {Haardt}  \& {Madau}}{{Volonteri}
  et~al.}{2003}]{Volonteri03}
{Volonteri} M.,  {Haardt} F.,   {Madau} P.,  2003, \mn@doi [\apj]
  {10.1086/344675}, \href {http://adsabs.harvard.edu/abs/2003ApJ...582..559V}
  {582, 559}

\bibitem[\protect\citeauthoryear{{Volonteri}, {Lodato}  \&
  {Natarajan}}{{Volonteri} et~al.}{2008}]{Volonteri08}
{Volonteri} M.,  {Lodato} G.,   {Natarajan} P.,  2008, \mn@doi [\mnras]
  {10.1111/j.1365-2966.2007.12589.x}, \href
  {http://esoads.eso.org/abs/2008MNRAS.383.1079V} {383, 1079}

\bibitem[\protect\citeauthoryear{{Volonteri}, {Silk}  \& {Dubus}}{{Volonteri}
  et~al.}{2015}]{VSD15}
{Volonteri} M.,  {Silk} J.,   {Dubus} G.,  2015, \mn@doi [\apj]
  {10.1088/0004-637X/804/2/148}, \href
  {http://adsabs.harvard.edu/abs/2015ApJ...804..148V} {804, 148}

\bibitem[\protect\citeauthoryear{{Volonteri}, {Reines}, {Atek}, {Stark}  \&
  {Trebitsch}}{{Volonteri} et~al.}{2017}]{Volonteri17}
{Volonteri} M.,  {Reines} A.,  {Atek} H.,  {Stark} D.~P.,   {Trebitsch} M.,
  2017, preprint, \href {http://adsabs.harvard.edu/abs/2017arXiv170400753V} {}
  (\mn@eprint {arXiv} {1704.00753})

\bibitem[\protect\citeauthoryear{{Wada}, {Papadopoulos}  \& {Spaans}}{{Wada}
  et~al.}{2009}]{Wada09}
{Wada} K.,  {Papadopoulos} P.~P.,   {Spaans} M.,  2009, \mn@doi [\apj]
  {10.1088/0004-637X/702/1/63}, \href
  {http://adsabs.harvard.edu/abs/2009ApJ...702...63W} {702, 63}

\bibitem[\protect\citeauthoryear{{Weigel}, {Schawinski}, {Treister}, {Urry},
  {Koss}  \& {Trakhtenbrot}}{{Weigel} et~al.}{2015}]{Weigel15}
{Weigel} A.~K.,  {Schawinski} K.,  {Treister} E.,  {Urry} C.~M.,  {Koss} M.,
  {Trakhtenbrot} B.,  2015, \mn@doi [\mnras] {10.1093/mnras/stv184}, \href
  {http://adsabs.harvard.edu/abs/2015MNRAS.448.3167W} {448, 3167}

\bibitem[\protect\citeauthoryear{{Wilkinson}}{{Wilkinson}}{1987}]{Wilkinson87}
{Wilkinson} D.~T.,  1987, in {Ulmer} M.~P.,  ed., 13th Texas Symposium on
  Relativistic Astrophysics. pp 209--218

\bibitem[\protect\citeauthoryear{{Willott}}{{Willott}}{2011}]{Willott11}
{Willott} C.~J.,  2011, \mn@doi [\apjl] {10.1088/2041-8205/742/1/L8}, \href
  {http://adsabs.harvard.edu/abs/2011ApJ...742L...8W} {742, L8}

\bibitem[\protect\citeauthoryear{{Yoshida}, {Omukai}  \& {Hernquist}}{{Yoshida}
  et~al.}{2008}]{Yoshida08}
{Yoshida} N.,  {Omukai} K.,   {Hernquist} L.,  2008, \mn@doi [Science]
  {10.1126/science.1160259}, \href
  {http://adsabs.harvard.edu/abs/2008Sci...321..669Y} {321, 669}

\bibitem[\protect\citeauthoryear{{Yue}, {Ferrara}, {Salvaterra}, {Xu}  \&
  {Chen}}{{Yue} et~al.}{2013}]{Yue13}
{Yue} B.,  {Ferrara} A.,  {Salvaterra} R.,  {Xu} Y.,   {Chen} X.,  2013,
  \mn@doi [\mnras] {10.1093/mnras/stt826}, \href
  {http://adsabs.harvard.edu/abs/2013MNRAS.433.1556Y} {433, 1556}

\bibitem[\protect\citeauthoryear{{Yue}, {Ferrara}, {Salvaterra}, {Xu}  \&
  {Chen}}{{Yue} et~al.}{2014}]{Yue14}
{Yue} B.,  {Ferrara} A.,  {Salvaterra} R.,  {Xu} Y.,   {Chen} X.,  2014,
  \mn@doi [\mnras] {10.1093/mnras/stu351}, \href
  {http://adsabs.harvard.edu/abs/2014MNRAS.440.1263Y} {440, 1263}

\bibitem[\protect\citeauthoryear{{de Bennassuti}, {Schneider}, {Valiante}  \&
  {Salvadori}}{{de Bennassuti} et~al.}{2014}]{deBen14}
{de Bennassuti} M.,  {Schneider} R.,  {Valiante} R.,   {Salvadori} S.,  2014,
  \mn@doi [\mnras] {10.1093/mnras/stu1962}, \href
  {http://adsabs.harvard.edu/abs/2014MNRAS.445.3039D} {445, 3039}

\bibitem[\protect\citeauthoryear{{de Bennassuti}, {Salvadori}, {Schneider},
  {Valiante}  \& {Omukai}}{{de Bennassuti} et~al.}{2017}]{deBennassuti17}
{de Bennassuti} M.,  {Salvadori} S.,  {Schneider} R.,  {Valiante} R.,
  {Omukai} K.,  2017, \mn@doi [\mnras] {10.1093/mnras/stw2687}, \href
  {http://adsabs.harvard.edu/abs/2017MNRAS.465..926D} {465, 926}

\makeatother
\end{thebibliography}

\appendix

\section{Properties of individual systems}
\label{appendix:prop}

In this appendix we provide a detailed description of 
the evolutionary histories of the four prototypical systems.

\begin{figure*}
\centering
\includegraphics [width=8.7cm]{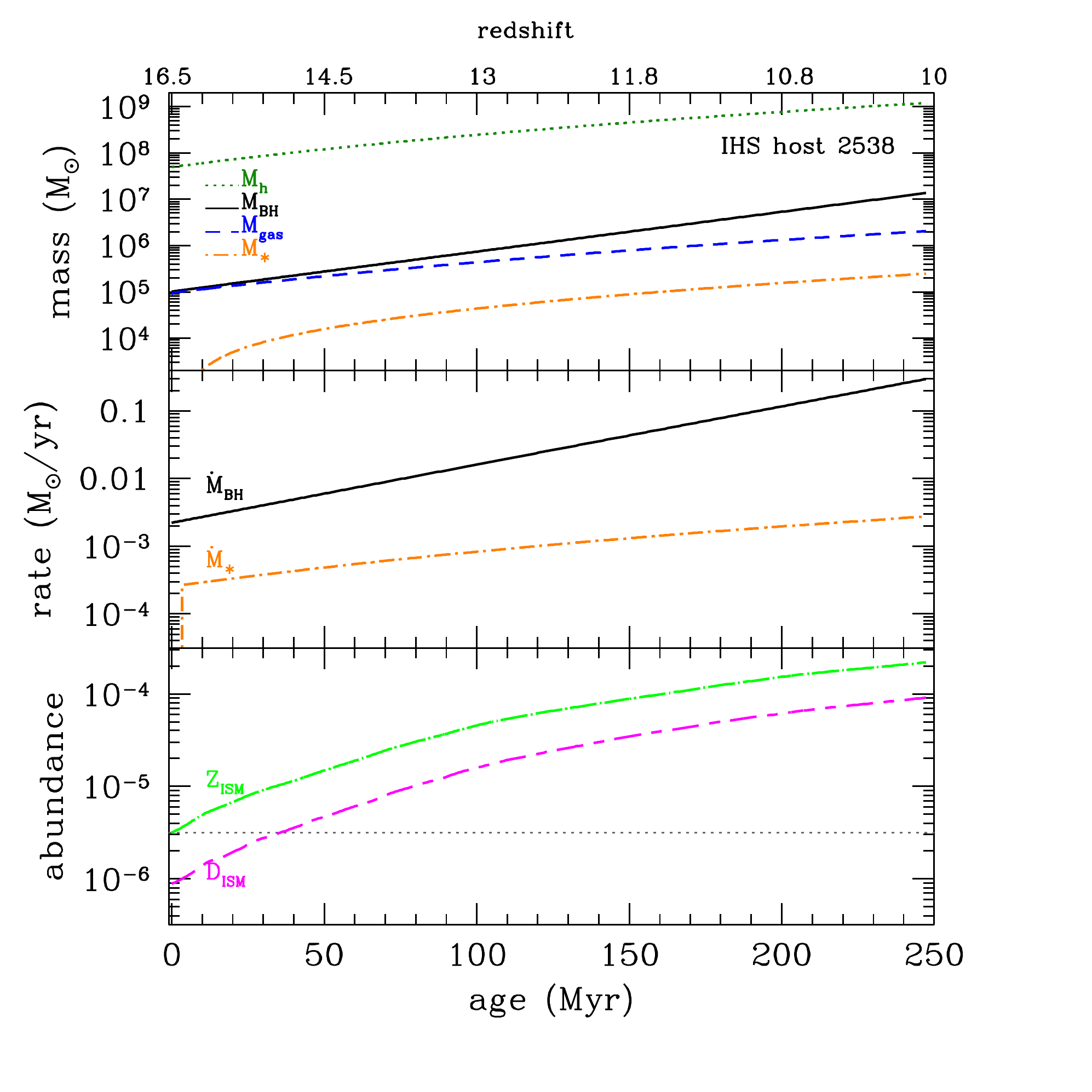}
\includegraphics [width=8.7cm]{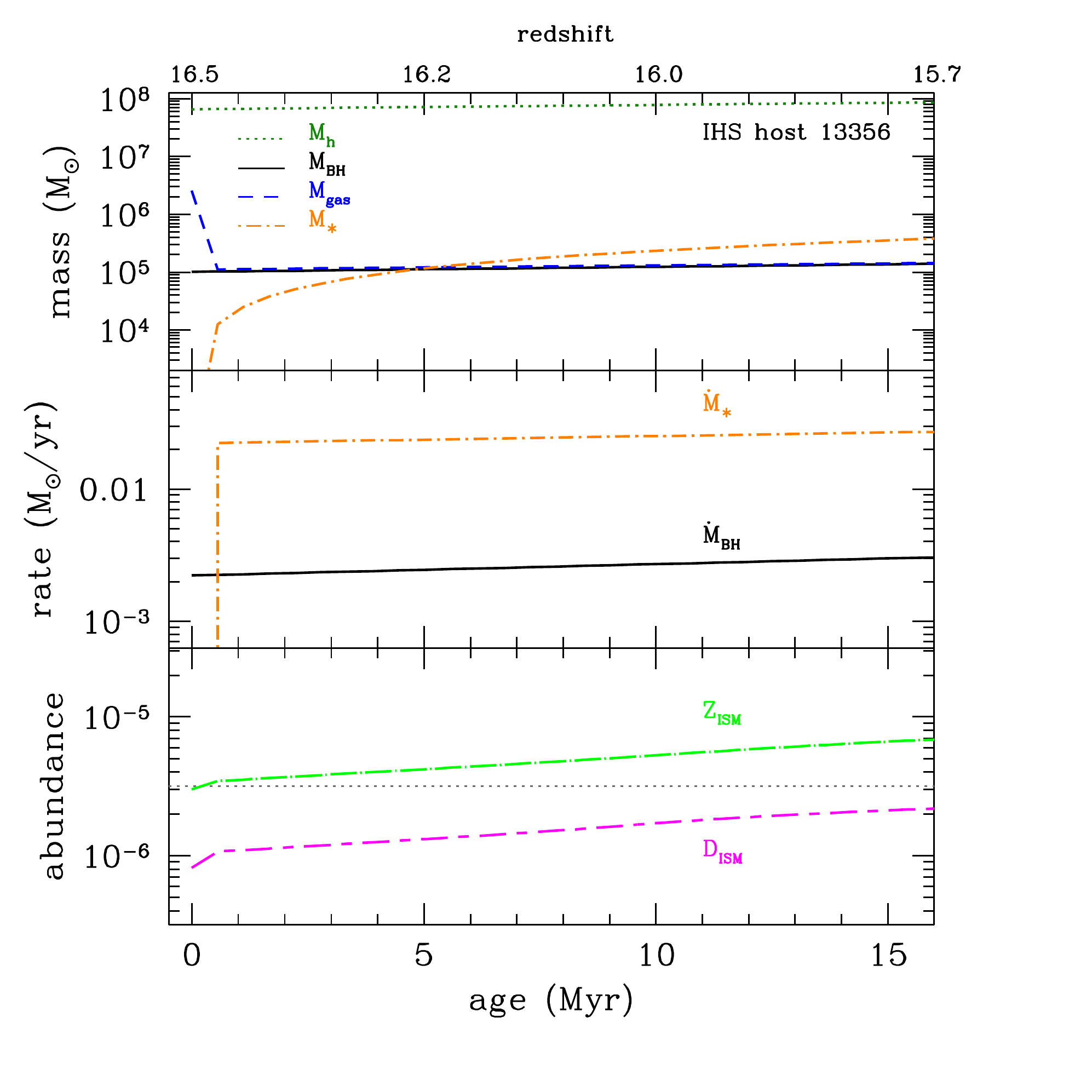}
\includegraphics [width=8.7cm]{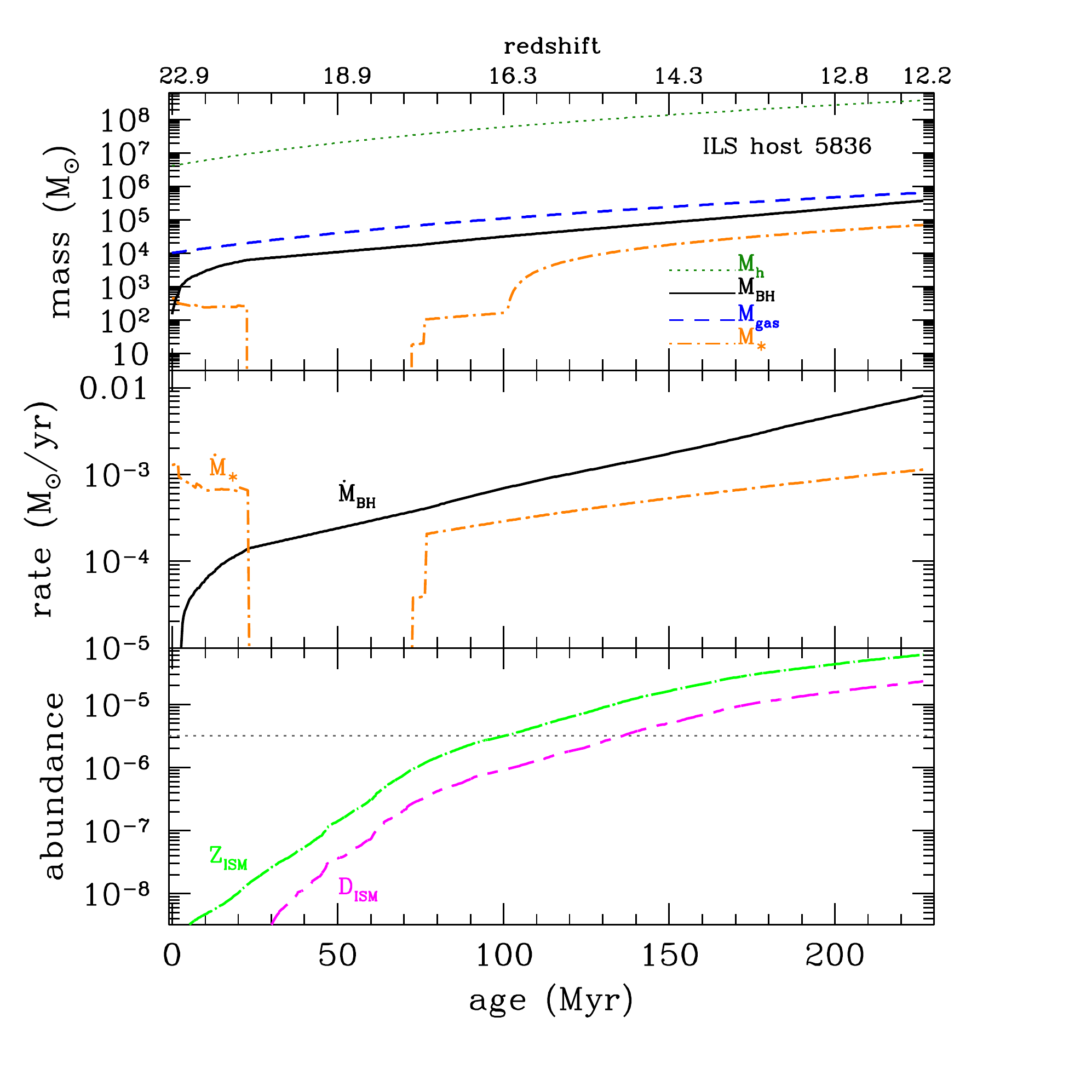}
\includegraphics [width=8.7cm]{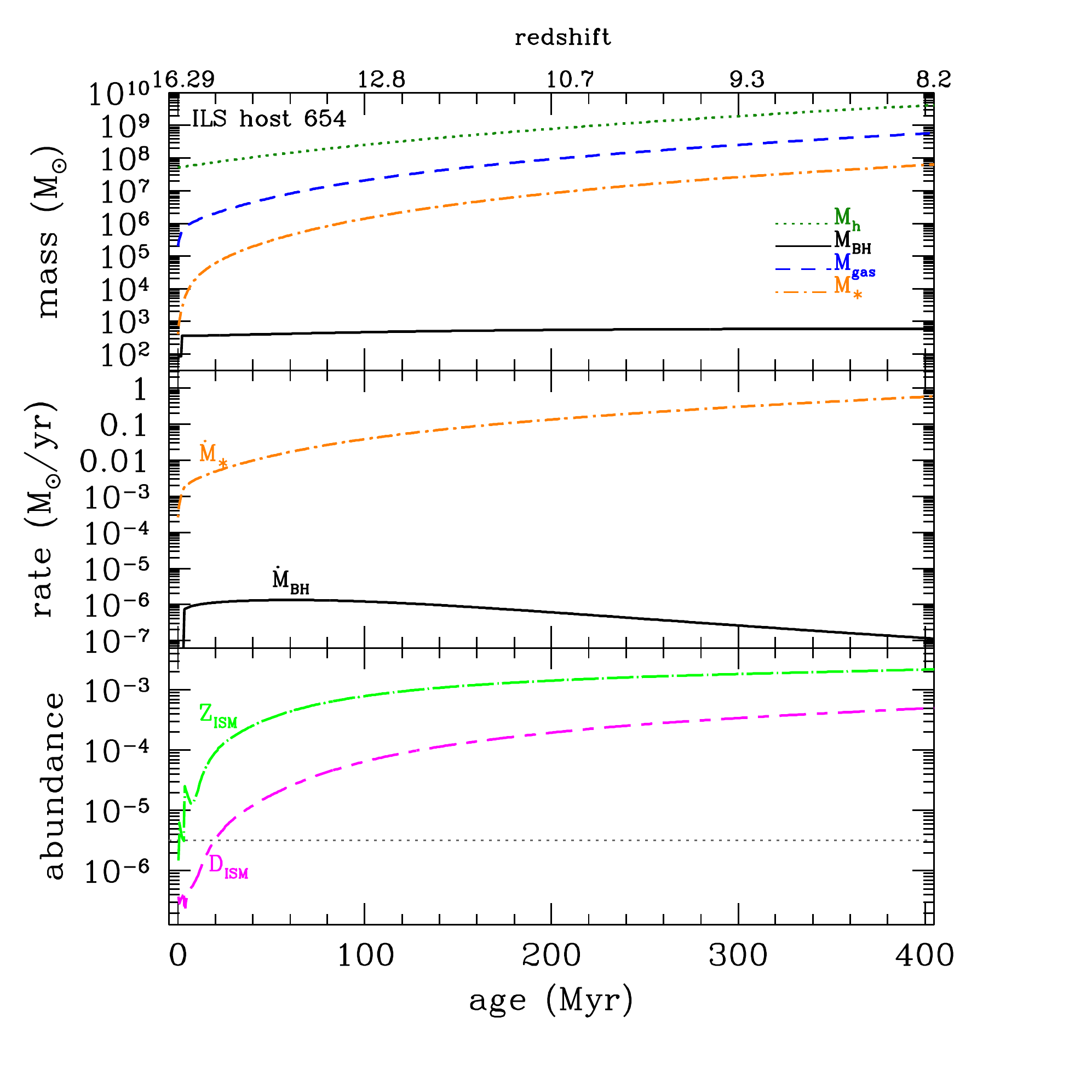}
\caption{Time dependent properties of IHS-2538 (top left), IHS-13356 (top right), ILS-5836 (bottom left)
and ILS-654 (bottom right). For each system, the upper panel shows the DM halo (dotted line), gas (dashed line),
stellar (dot-dashed line) and BH (solid line) masses. The middle panel shows the star formation (dot-dashed)
and black hole accretion (solid) rates. The lower panel shows the metallicity (dotted-dashed line) and dust-to-gas mass
ratio (dashed line) of the ISM, 
in absolute units. The horizontal dotted line indicates the critical metallicity for the Pop~III/Pop~II star transition.} 
\label{fig:prop} 
\end{figure*}

We present the time dependent properties of
systems IHS-2538, IHS-13356, ILS-5836 and ILS-654 in Fig.~\ref{fig:prop}.
For each system, in the upper panel we show the DM (dotted line), gas (dashed),
stellar (dot-dashed) and BH (solid) masses. 
In the upper horizontal axis we also label the redshifts in the simulation that correspond
to the different ages.
The star formation and BH accretion rates are shown in the middle panel
with dot-dashed and solid lines respectively. 
Finally, the lower panel reports the gas-phase metals ($Z_{\rm ISM}$, dotted-dashed line) 
and dust-to-gas mass ratio ($D_{\rm ISM}$, dashed line) in the ISM (in absolute units).
For comparison, we also show the critical metallicity, $Z_{\rm cr}\sim 10^{-4} \, Z_\odot$ 
(dotted horizontal line), which marks the transition from Pop~III to Pop~II star formation.
%

\noindent \textit{IHS-2538}: this is one of the most massive and longest-living isolated heavy seeds.
By construction, the initial BH has a mass of $10^5 \, M_\odot$ and forms in a 
$\rm Ly\alpha$ cooling halo ($M_{\rm h} \sim 5\times 10^7 \,  M_\odot$) at redshift $z=16.5$.
Such a DM halo is the descendant of two sterile mini-halos. For its entire history, the heavy seed grows 
close to or at the Eddington rate. Efficient BH feeding and feedback enable the galaxy to sustain only a 
very low star formation rate ($< 3 \times 10^{-3} \, M_\odot/{\rm yr}$). 
As a result, the BH always grows faster than the stellar component, and the SED properties are expected 
to be dominated by the AGN emission. \\

\noindent \textit{IHS-13356}: this is one of the shortest-living and smallest isolated heavy seeds.
Differences in the histories of IHS-13356 progenitors, relative to IHS-2538, 
lead to different initial gas masses at $\rm age=0$. 
The reason for this behaviour can be understood as follows:
newly virialized halos gradually accrete their gas mass following an exponentially 
decaying infall rate until they merge with another halo
\citep[i.e. they do not instantaneously acquire the 
universal baryon fraction $\Omega_{\rm b}/\Omega_{\rm M}$; see eq.~3 and furter 
details in][]{SS08}.
IHS-13356 is hosted in a $\rm Ly\alpha$ cooling halo
formed by the merger of two sterile mini-halos at $z = 16.5$. These ancestors virialize
at $z \sim 17$ and accrete a gas mass which corresponds to $\sim 40\%$ of $\Omega_{\rm b}/\Omega_{\rm M}$
prior to their coalescence. Conversely, IHS-2538 is hosted in a 
$\rm Ly\alpha$ cooling halo that has formed soon after its progenitor mini-halos virialize, so that
they accrete only $\sim 1\%$ of $\Omega_{\rm b}/\Omega_{\rm M}$ by the time they coalesce.
When IHS-13356 forms, the coalescence of its gas-rich ancestors triggers an efficient starburst of 
$\sim 2\times 10^{-2} \, M_\odot/{\rm yr}$.
The gas mass suddenly drops from $\sim 2 \times 10^6 \, M_\odot$ to 
$\sim 10^5 \, M_\odot$ within the first 0.5 Myr, corresponding to the time step of the simulation
at that redshift. Thereafter, IHS-13356 evolves in gas starvation, with star formation, BH accretion
and the associated mechanical feedback depleting almost all the gas mass that the halo is accreting
from the intergalactic medium. At the end of the brief (16 Myr) isolated evolution, the nuclear BH
mass has increased by $\sim 50 \%$ and the stellar mass is a factor $\sim 2.5$ larger. 
The emission properties of this system are dominated by the AGN during the first 5 Myr 
of the evolution (when $M_\ast < 10^5 \, M_\odot$), and by the emission of the host galaxy thereafter.\\

\noindent \textit{ILS-5836}: this is one of the most massive isolated light seeds.
Its host halo is one of the first star forming mini-halos ($M_{\rm h} \sim 4 \times 10^6 \, M_\odot$) 
at redshift $z \sim 23$.
Due to its pristine composition, a first burst of Pop~III stars ($10^{-3} \, M_\odot/{\rm yr}$) leads 
to the formation of one massive BH remnant with $M_{\rm BH} \sim 70 \, M_\odot$.
Pop~III star formation continues for about 100 Myrs, down to redshift $z \sim 16.2$, but with 
a $\sim 50$~Myr suppression due to the effect of the LW flux at which the mini-halo is 
exposed. Due to their short evolutionary timescales, massive Pop~III stars are assumed to 
evolve in one simulation timestep (instantaneous recycling approximation). 
Hence $M_\ast=0$ when the $\dot{M}_\ast=0$.
When the DM halo has become a $\rm Ly\alpha$ cooling halo, Pop~III star formation turns on again and 
continues until metal enrichment triggers the transition to Pop~II star formation, in the last 
$\sim 125$~Myr of evolution.
During the Pop~III regime, the nuclear BH grows by gas accretion and mergers with other remnant BHs
forming in subsequent bursts, reaching a mass of $\sim 10^4 \, M_\odot $ in $\sim 70$ Myr.
The gas accretion rate triggered during the non-star forming phase leads to a BH dominance 
scenario even in the subsequent Pop~II star formation regime: in ILS-5836, the BH mass continues 
to grow faster than its stellar counterpart. 
At $z \sim 12.2$, the BH mass has grown to $4 \times 10^5 \, M_\odot$. 
This evolution is reflected in the SED, that is a sort of
scaled-down version of the SED of IHS-2538.\\

\noindent \textit{ILS-654}: this is one of the least massive but longest-living isolated light seeds.
Inefficient BH accretion [$\dot{M}_{\rm BH}\sim (10^{-7}-10^{-6}) \, M_\odot/{\rm yr}$] 
leads to a final mass of only $800 \, M_\odot$ in more than 400 Myr of evolution.
Although they have similar initial BH masses and accretion rates, $\sim 10^{-6} \, M_{\odot}/{\rm yr}$ at 
$\rm age=0$, BH-growth in ILS-5836 at early times is mainly driven by mergers with other Pop~III BH 
remnants during the prolonged Pop~III star formation in this halo.
As a consequence of the low BH growth of ILS-654, the available gas mass can efficiently fuel star 
formation, yielding to a star formation rate $> 0.01 \, M_\odot/{\rm yr}$ already at $\rm age>30$~Myr.
The evolution of ILS-654 is indeed consistent with that of a normal star-forming galaxy, and the system 
becomes the most metal-rich of the four selected objects. In addition, in the absence of efficient BH 
accretion and feedback more infalling gas is retained and the higher metallicity, dust-to-gas ratio and 
gas density are reflected in the predicted SED.

\bsp
\label{lastpage}
\end{document}